\documentclass[fleqn,usenatbib]{mnras}

\usepackage{newtxtext,newtxmath}
\usepackage[T1]{fontenc}
\usepackage{graphicx}	% Including figure files
\usepackage{amsmath}	% Advanced maths commands
\usepackage{amssymb}	% Extra maths symbols
\usepackage{mathtools}                                             \usepackage{subfigure}  
\usepackage{color}                                                  
\usepackage{xcolor}                                                 
\usepackage{hyperref}                                                  

\newcommand{\dif}{\mathrm{d}}
\newcommand{\HI}{\rm H\,{\sc I}}

\newcommand{\NSFPDF}{NSF Astronomy and Astrophysics Postdoctoral Fellow}
\newcommand{\JANPDF}{Nithyanandan Thyagarajan is a Jansky fellow of the National Radio Astronomy Observatory}

\graphicspath{{./}{figures/}}

\date{}
\pubyear{2020}
\title[GPR foreground modelling in HERA-47]{Foreground modelling via Gaussian process regression: an application to HERA data}

\author[A. Ghosh et al.]{
Abhik Ghosh,$^{1,2,3}$\thanks{E-mail: aghosh@ska.ac.za}
Florent Mertens,$^{4,5}$
Gianni Bernardi,$^{7,8,2}$
M\'ario G. Santos,$^{1,2}$
\newauthor
Nicholas S. Kern,$^{9}$
Christopher L. Carilli,$^{10,11}$
Trienko L. Grobler,$^{12,8}$
L\'eon V. E. Koopmans,$^{4}$
\newauthor
Daniel C. Jacobs,$^{13}$
Adrian Liu,$^{14,9}$
Aaron R. Parsons,$^{9}$
Miguel F. Morales,$^{15}$
\newauthor
James E. Aguirre,$^{16}$
Joshua S. Dillon,$^{9}$\thanks{\NSFPDF}
Bryna J. Hazelton,$^{15,17}$
Oleg M. Smirnov,$^{2,8}$
\newauthor
Bharat K. Gehlot,$^{4,13}$
Siyanda Matika,$^{8}$
Paul Alexander,$^{11}$
Zaki S. Ali,$^{9}$
Adam P. Beardsley,$^{13}$\thanks{\NSFPDF}
\newauthor
Roshan K. Benefo,$^{18}$
Tashalee S. Billings,$^{16}$
Judd D. Bowman,$^{13}$
Richard F. Bradley,$^{19}$
\newauthor
Carina Cheng,$^{9}$
Paul M. Chichura,$^{16}$
David R. DeBoer,$^{9}$
Eloy de~Lera~Acedo,$^{11}$
\newauthor
Aaron Ewall-Wice,$^{20}$
Gcobisa Fadana,$^{2}$
Nicolas Fagnoni,$^{11}$
Austin F. Fortino,$^{16}$
Randall Fritz,$^{2}$
\newauthor
Steve R. Furlanetto,$^{21}$
Samavarti Gallardo,$^{16,22}$
Brian Glendenning,$^{10}$
Deepthi Gorthi,$^{9}$
\newauthor
Bradley Greig,$^{23,24}$
Jasper Grobbelaar,$^{2}$
Jack Hickish,$^{9}$
Alec Josaitis,$^{11}$
Austin Julius,$^{2}$
\newauthor
Amy S. Igarashi,$^{18,25}$
MacCalvin Kariseb,$^{2}$
Saul A. Kohn,$^{16}$
Matthew Kolopanis,$^{13}$
\newauthor
Telalo Lekalake,$^{2}$
Anita Loots,$^{2}$
David MacMahon,$^{9}$
Lourence Malan,$^{2}$
Cresshim Malgas,$^{2}$
\newauthor
Matthys Maree,$^{2}$
Zachary E. Martinot,$^{16}$
Nathan Mathison,$^{2}$
Eunice Matsetela,$^{2}$
\newauthor
Andrei Mesinger,$^{26}$
Abraham R. Neben,$^{20}$
Bojan Nikolic,$^{11}$
Chuneeta D. Nunhokee,$^{8,16}$
\newauthor
Nipanjana Patra,$^{9}$
Samantha Pieterse,$^{2}$
Nima Razavi-Ghods,$^{11}$
Jon Ringuette,$^{16}$
James Robnett,$^{10}$
\newauthor
Kathryn Rosie,$^{2}$
Raddwine Sell,$^{2}$
Craig Smith,$^{2}$
Angelo Syce,$^{2}$
Max Tegmark,$^{20}$
\newauthor
Nithyanandan Thyagarajan,$^{10,13}$\thanks{\JANPDF}
Peter K.~G. Williams,$^{27,28}$
Haoxuan Zheng$^{20}$
\\
$^{1}$Department of Physics and Astronomy, University of Western Cape, Cape Town 7535, South Africa\\
$^{2}$The South African Radio Astronomy Observatory (SARAO), 2 Fir Street, Black River Park, Observatory, 7925, South Africa\\
$^{3}$Department of Physics, Banwarilal Bhalotia College, GT Rd, Ushagram, Asansol, West Bengal, India\\
$^{4}$Kapteyn Astronomical Institute, University of Groningen, P. O. Box 800, 9700 AV Groningen, The Netherlands\\
$^{5}$LERMA, Observatoire de Paris, PSL Research University, CNRS, Sorbonne Université, F-75014 Paris, France\\
$^{7}$INAF - IRA, via P. Gobetti 101, I-40129 Bologna, Italy\\
$^{8}$Department of Physics and Electronics, Rhodes University, PO Box 94, Grahamstown 6140, South Africa\\
$^{9}$Department of Astronomy, University of California, Berkeley, CA\\
$^{10}$National Radio Astronomy Observatory, Socorro, NM\\
$^{11}$Cavendish Astrophysics, University of Cambridge, Cambridge, UK\\
$^{12}$Dept of Mathematical Sciences, Computer Science Division, Stellenbosch University, Private Bag X1, 7602 Matieland, South Africa\\
$^{13}$School of Earth and Space Exploration, Arizona State University, Tempe, AZ\\
$^{14}$Department of Physics and McGill Space Institute, McGill University, 3600 University Street, Montreal, QC H3A 2T8, Canada\\
$^{15}$Department of Physics, University of Washington, Seattle, WA\\
$^{16}$Department of Physics and Astronomy, University of Pennsylvania, Philadelphia, PA\\
$^{17}$eScience Institute, University of Washington, Seattle, WA\\
$^{18}$Center for Particle Cosmology, Department of Physics and Astronomy, University of Pennsylvania, Philadelphia, PA 19104 USA\\
$^{19}$National Radio Astronomy Observatory, Charlottesville, VA\\
$^{20}$Department of Physics, Massachusetts Institute of Technology, Cambridge, MA\\
$^{21}$Department of Physics and Astronomy, University of California, Los Angeles, CA\\
$^{22}$California State University of Los Angeles, 5151 State University Dr, Los Angeles, CA 90032 USA\\
$^{23}$School of Physics, University of Melbourne, Parkville, VIC 3010, Australia\\
$^{24}$ARC Centre of Excellence for All-Sky Astrophysics in 3 Dimensions (ASTRO 3D), University of Melbourne, VIC 3010, Australia\\
$^{25}$Department of Astronomy, San Diego State University, San Diego, CA 92182 USA\\
$^{26}$Scuola Normale Superiore, 56126 Pisa, PI, Italy\\
$^{27}$Center for Astrophysics | Harvard \& Smithsonian, Cambridge, MA\\
$^{28}$American Astronomical Society, Washington, DC
}
\begin{document}
\label{firstpage}
\pagerange{\pageref{firstpage}--\pageref{lastpage}}
\maketitle

\begin{abstract}
The key challenge in the observation of the redshifted 21-cm signal from cosmic reionization is its separation from the much brighter foreground emission.  Such separation relies on the different spectral properties of the two components, although, in real life, the foreground intrinsic spectrum is often corrupted by the instrumental response, inducing systematic effects that can further jeopardize the measurement of the 21-cm signal. In this paper, we use Gaussian Process Regression to model both foreground emission and instrumental systematics in $\sim 2$~hours of data from the Hydrogen Epoch of Reionization Array.
We find that a simple co-variance model with three components matches the data well, giving a residual power spectrum with white noise properties.
These consist of an ``intrinsic" and instrumentally corrupted component with a coherence-scale of 20~MHz and 2.4~MHz respectively (dominating the line of sight power spectrum over scales $k_{\parallel} \le 0.2$~h~cMpc$^{-1}$) and a baseline dependent periodic signal with a period of $\sim 1$~MHz (dominating over $k_{\parallel} \sim 0.4 - 0.8$~h~cMpc$^{-1}$) which should be distinguishable from the 21-cm EoR signal whose typical coherence-scales is $\sim 0.8$~MHz.
\end{abstract}


\section{Introduction} \label{sec:intro}
Observations of the redshifted 21-cm signal from neutral Hydrogen hold the promise of revealing the detailed astrophysical processes occurring during the Epoch of Reionization (EoR) and the Cosmic Dawn (CD). The 21-cm signal can provide insights into the formation and evolution of the first structures in the Universe~\citep[see, e.g.,][for reviews]{Furlanetto06,Morales10,Pritchard12,Mellema13}: for example, when the intergalactic medium (IGM) is still largely neutral, it is a sensitive probe of the first sources of Lyman-$\alpha$ and X-ray radiation \citep{Mesinger07, Santos10, Santos11, McQuinn12,Fialkov14,Fialkov17} and, during the subsequent EoR, its large-scale fluctuations map the evolution of the global ionization fraction \citep{Lidz08,BoltonXHI11}. The 21-cm emission gives insights into the nature of formation of the first stars, galaxies and their impact on the physics of the IGM \citep{Loeb13, Zaroubi13}.

At present, several experiments are attempting to detect the power spectrum of the 21-cm signal from the EoR (e.g. GMRT\footnote{\url{http://www.gmrt.ncra.tifr.res.in/}}, LOFAR\footnote{Low Frequency Array, \url{http://www.lofar.org}}, MWA\footnote{Murchison Widefield Array, \url{http://www.mwatelescope.org}}, PAPER\footnote{Precision Array to Probe EoR, \url{http://eor.berkeley.edu}}) or the sky-averaged 21-cm emission using a single dipole \citep{Bowman10, Patra15, Greenhill12, Bernardi16}. Some of these ongoing efforts have achieved increasingly better upper limits on the 21-cm signal power spectra~\citep{Li19,Barry19,Kolopanis19,Patil17,Beardsley16,Ali15}, showing the way for the second generation experiments such as the Square Kilometre Array (SKA\footnote{\url{http://www.skatelescope.org}}) and the Hydrogen Epoch of Reionization Array (HERA\footnote{\url{http://reionization.org}}). Recently, a detection of an absorption profile in the sky-averaged 21-cm signal centred at 78~MHz has been reported \citep{Bowman18}, although the unexpected depth of the trough is calling for independent confirmations \citep{Fraser18} - including interferometric observations \citep{Gehlot19}.

The main challenge in detecting the faint 21-cm signal is the presence of Galactic and extra-galactic foregrounds that are around 3-4 orders of magnitude stronger~\citep[e.g.][]{Bernardi09,Bernardi10,Ghosh11b,Dillon14,Parsons14}. Foregrounds as well as the instrumental response have a highly-correlated continuum spectrum and can, in principle, be separate from the 21-cm signal that has structure on smaller frequency scales due to the intrinsic redshift evolution of the IGM \citep[e.g.,][]{Bharadwaj01, Zaldarriaga04, Santos05}. 
However, the inherent smoothness of the foreground emission is often compounded by the interferometric response (``mode-mixing"), including frequency-dependent primary beams, side-lobe from bright, mis-subtracted sources and ionospheric distortions~\citep{Bowman09,Koopmans10,Ghosh11b,Vedantham12,Yatawatta13,Vedantham16,Barry16,Patil17,Gehlot19,Byrne19}. Polarization leakage due to improper calibration may also add additional spectral structures to the unpolarized cosmological 21-cm window \citep{Geil11,Asad15,Nunhokee17}.  

The study of foreground properties and their separation from the 21-cm signal have been a very active research area over the years \citep[e.g.,][]{Datta10, Liu11, Trott12, Morales12, Dillon14}. One strategy is to attempt to ``avoid" foregrounds, i.e. to avoid $k$ modes which are contaminated by foregrounds and to estimate the 21-cm power spectrum using the uncontaminated modes. This assumes that foregrounds are well localized in $k$-space and the mode-mixing effects can be kept well under control \citep{Thyagarajan15}. This foreground avoidance method has the disadvantage of considerably reducing the sensitivity of the instrument, because of reduction in the number of $k$-modes that can be probed to characterize the EoR signal \citep{Pober14}. The second approach involves subtracting the best possible foreground model and, possibly, recover access to the foreground dominated power spectrum region. One of the possible disadvantages here is the risk of contamination of the cosmological 21-cm signal from the cleaning process. Foreground wedge also corrupts nearly all the redshift space 21-cm signal, making it difficult to extract cosmological information without foreground subtraction \citep{Pober15,Jensen16}. There are also recent efforts to develop a somewhat hybrid analysis where a GLEAM \citep[Galactic and Extragalactic All-sky MWA;][]{Hurley17} catalog of sources including Pictor A and Fornax A were first subtracted from PAPER-64 (Precision Array for Probing the Epoch of Re-ionisation) data and then the power spectrum was estimated. This is equivalent to an additional visibility-based filtering within the foreground avoidance paradigm \citep{Kerrigan18}.

\citet{Chapman14} pointed out that blind foreground removal methods such as Generalized Morphological Component Analysis \citep[GMCA,][]{Chapman13} can still model relatively non-smooth foregrounds effectively on short baselines ($k_{\perp} \lesssim 0.2$~h~cMpc$^{-1}$), while avoidance suffers some degradation as the frequency-dependent small-scale structure cannot be confined purely in a region at small $k_{\parallel}$ modes. 
Several techniques have been proposed to model and remove foreground emission taking advantage of its spectral smoothness, including parametric \citep{Jelic08,Bonaldi15} and non-parametric methods \citep{Harker09,Chapman13, Mertens18, Mertens20}. 
Both methods have the limitation that they may suppress the 21-cm signal and do not always reach a level of modeling error better than the noise for $k \le 0.3$~h~cMpc$^{-1}$, compared to the desired level of the 21-cm signal power spectrum \citep{Mertens18}. In general, foreground subtraction allows to use virtually all $k$ modes at the risk of contamination of the 21-cm signal, whereas foreground avoidance does not corrupt the cosmological signal within the EoR window, but can not take advantage of any mode in the foreground wedge.

Recently, a novel, non-parametric (in the signal) method based on Gaussian Process Regression (GPR) has been studied in detail with simulations, where intrinsic smooth foregrounds, mid-scale frequency fluctuations associated with mode-mixing, Gaussian random noise, and a basic 21-cm signal model, are modelled with Gaussian Process (GP), and subsequently a separation with a precise estimation of the uncertainty was carried out \citep{Mertens18, Gehlot19}. The advantage of this method over previous ones is its implementation in a Bayesian framework that allows to incorporate different physical processes in the form of covariance structure priors (currently spectral and possible spatial implementation in future) on the various components. GPR further allows much better control over the coherence structure (and hence power spectra) of all components rather then be ``blind” for their physical origins, as are Generalized Morphological Component Analysis (GMCA), Independent Component Analysis (ICA), or fitting polynomials. Further, it also offers a good way to extract foreground models from the data.

%{\bf This gives us access to the foreground dominated power spectrum region and thereby enables access to the large scale line of sight modes. These modes have the highest SNRs and hence increases the sensitivity of 21 cm analyses.}

In this paper, we apply GPR to model foregrounds in a $\sim 100$~minute long observation with HERA-47. The GPR method was originally developed to be applied to observations with good $uv$~coverage, but here we adapted it to work directly to visibilities, without being affected by the sparse HERA $uv$~coverage. Foreground modeling helps us to assess the level of contamination of the data and the covariance models that can properly describe foregrounds. It can ultimately guide the foreground cleaning process and help finding the scales which should be safe to use in a foreground avoidance approach. We use the line of sight and the delay power spectrum in the $(k_{\perp},k_{\parallel})$ plane as our metric to characterize the foreground models. 

The paper is organized as follows: Section~\ref{sec:hera-data} summarizes our observations along with the delay power spectrum estimation procedure, Section~\ref{sec:gpr} describes our technique to calculate the foreground power spectrum using the GPR formalism. Finally, we conclude in Section~\ref{sec:discussion}.
Cosmological parameters used here are from \citet{PLANCK16}.

\section{Observations and data reduction} \label{sec:hera-data}
The Hydrogen Epoch of Reionization Array \citep{Deboer17} is an ongoing experiment to use the red-shifted 21-cm radiation originating from the cosmological distribution of neutral hydrogen ($\HI$) to study the formation of first stars and black holes from CD ($z \sim 30$) to the full IGM ionization history ($6 \lesssim z \lesssim 12$). In its final configuration, the array will consist of 350 parabolic dishes of $\sim 14$~m diameter, with an effective area of $\sim 154$~m$^2$ per antenna, closely packed in a hexagonal split-core \citep{Dillon16}, plus outriggers up to $\sim 1$~km distance. The experiment is optimized for robust power spectrum detection while minimizing foreground contamination \citep{Pober14,Ewall-Wice17,Thyagarajan15}.

\begin{figure}
\includegraphics[width=80mm,angle=0]{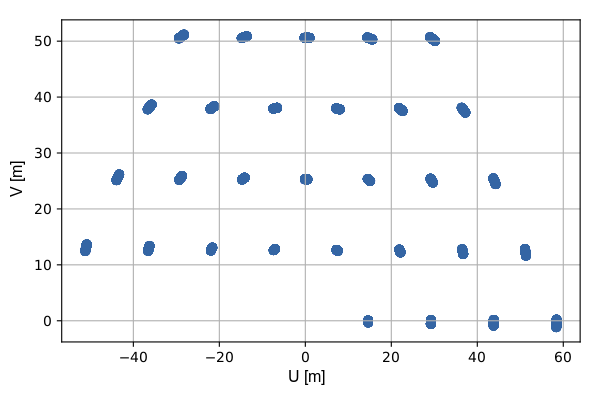}
\caption{This figure shows the 10~minute $uv$ coverage around 150 MHz of HERA-47 which are analyzed in this paper.}
\label{fig:uv_coverage}
\end{figure}

In this paper, we used data from the deployment of the first 47 dish (HERA--47) array. It covers a frequency range of $100 - 200$~MHz with a channel resolution of $\sim 97.6$~kHz. The results presented in this paper were generated from ten nights of HERA--47 data, starting on 2017-10-15, using only the `xx' polarization cross-products. In the paper, we refer to this as stokes `I'. We selected snapshots of 10 minutes (see Figure~\ref{fig:uv_coverage} for the corresponding $uv$~coverage) of data close to $21^{\rm h}$ LST over 10 days from the HERA data repository\footnote{\url{https://github.com/HERA-Team/librarian/}}. In total, we used around 100 minutes of data. We used the \textsc{pyuvdata}~\footnote{\mbox{\url{https://github.com/RadioAstronomySoftwareGroup/pyuvdata}}} software \citep{Hazelton17} to convert the correlator output to the Common Astronomy Software Applications (\textsc{CASA})\footnote{\url{http://casa.nrao.edu}} Measurement Set format. Antenna 50 was found to be bad for the initial seven days and was permanently flagged. We also flagged the band edges as well as the channels that were persistently affected by Radio Frequency Interference (RFI), i.e. mostly the following channels: 0--100, 379--387, 510--512, 768--770, 851--852 and 901--1023, where channel `0' corresponds to 100 MHz and channel `1023' to 200 MHz. We then made use of the \textsc{CASA} task \textsc{rflag} to perform further flagging in time and frequency. The threshold for `timedevscale' and `freqdevscale' was fixed to the default values of `5' each. This implies that for each channel any visibility will be flagged if the local RMS of its real and imaginary part, is larger than $5$ times  (RMS + median deviation) within a sliding time window. Similarly, for each integration time, the real and imaginary parts of the visibilities were flagged if they exceed $5$ times the deviation from the median value across channels.

Calibration was performed using custom CASA pipelines \footnote{\url{https://github.com/Trienko/heracommissioning}}. The starting flux density model included the five brightest point sources within the HERA field of view (GLEAM 2101-2800, GLEAM 2107-2525, GLEAM 2107-2529, GLEAM 2101-2803 and GLEAM 2100-2829), chosen from the MWA GLEAM point source catalog \citep{Hurley17}.
Their model flux density was corrected for the HERA primary beam response following the electromagnetic simulations of the HERA feed and dish \citep{Fagnoni16}, obtaining a flux density estimate for each source at each frequency channel. This sky model was used to solve for three types of antenna gains: antenna-based delay (`K' term in the CASA terminology), followed by a complex gain for all the channels and the whole 10~minute interval (`G' term in the CASA terminology) and by a complex bandpass calibration (`B' term in the CASA terminology). Calibration solutions were determined for the snapshot observation on 2017-10-15 and applied to the rest of the nine nights of data, though data from each day was flagged individually.
The calibration solutions are used as-is, and are not smoothed across frequency before applying them to the data. This can allow spectrally-dependent calibration errors generated by unmodelled sky sources and baseline-dependent systematics to be applied to the data, and can further corrupt the EoR window \citep{Kern19c}; however, in this work we seek to model these terms through a combination of the foreground mode mixing and periodic kernel discussed below.

Calibrated visibilities were phased to a common right ascension $\alpha = 21^{\rm h}$ and Fourier transformed into $21^\circ \times 21^\circ$ images using the w-projection algorithm with 128 planes and the multi-frequency synthesis algorithm to combine the whole bandwidth together. Uniform weights were used, leading to a $43.2' \times 35.4'$ synthesized beam. Each image was conservatively deconvolved down to a threshold of $10\%$ of the image peak using the Cotton-Schwab algorithm implemented in the CASA {\it CLEAN} task.

\begin{figure*}
\includegraphics[width=55mm,angle=0]{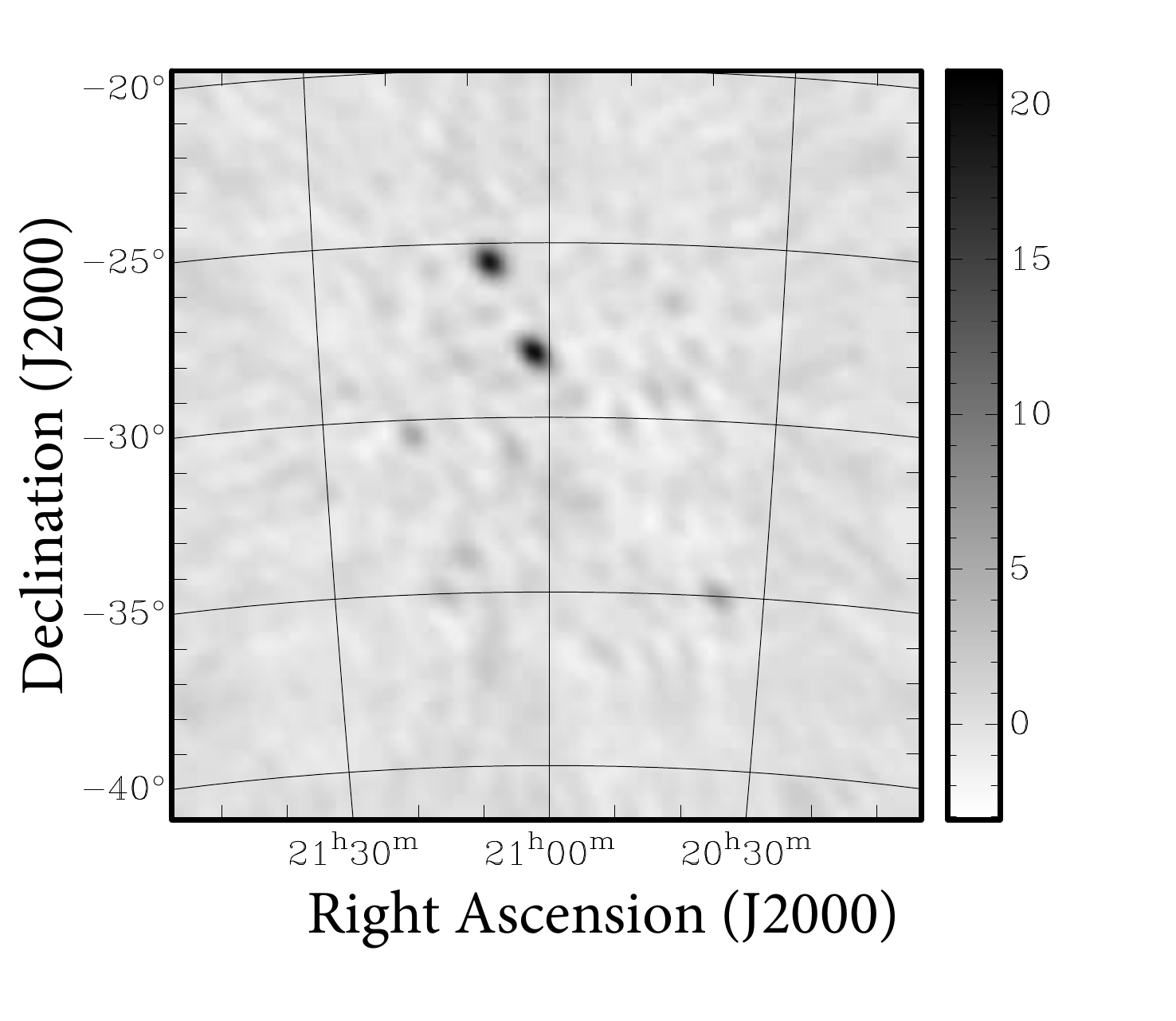}
\includegraphics[width=55mm,angle=0]{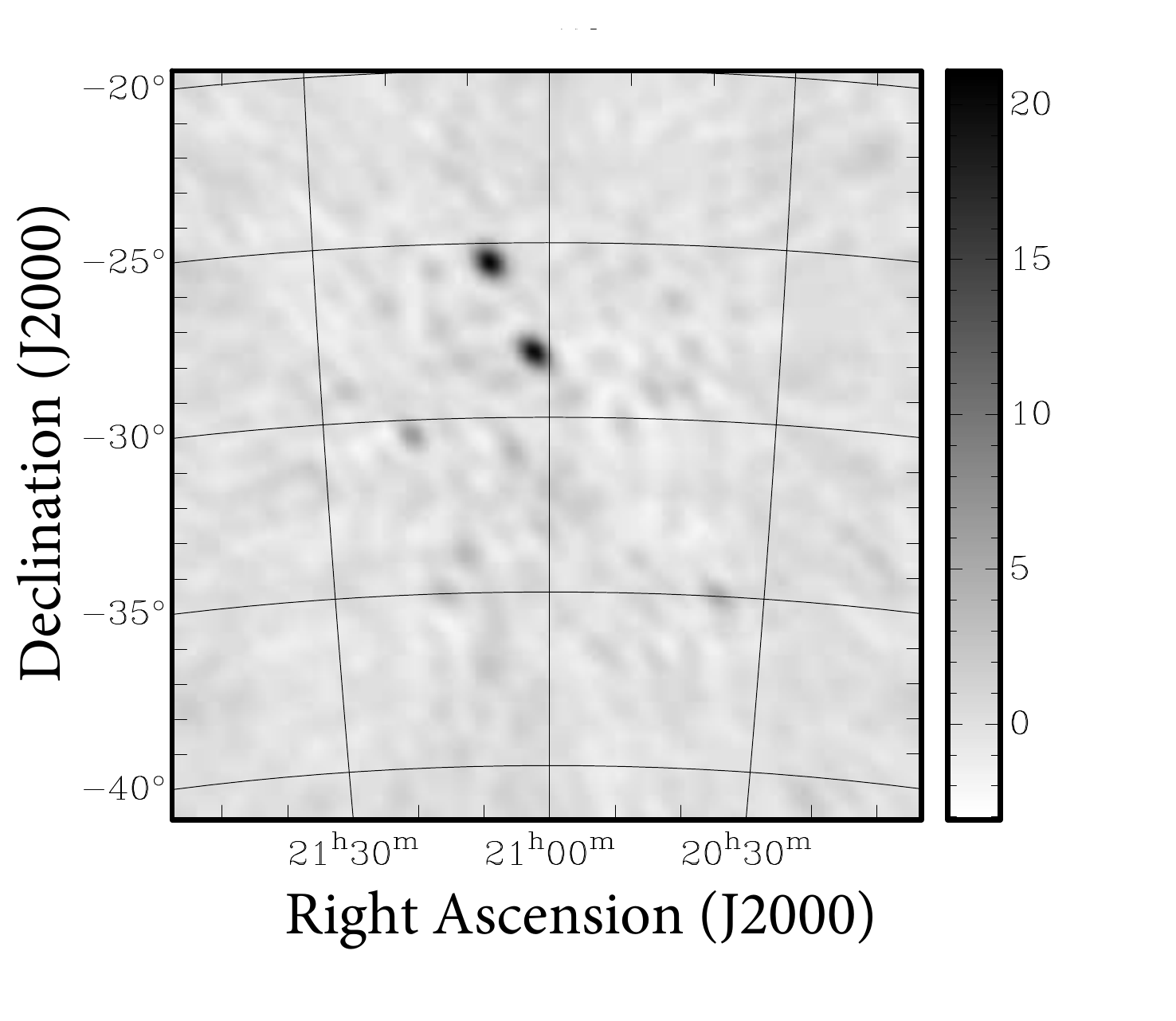}
\includegraphics[width=55mm,angle=0]{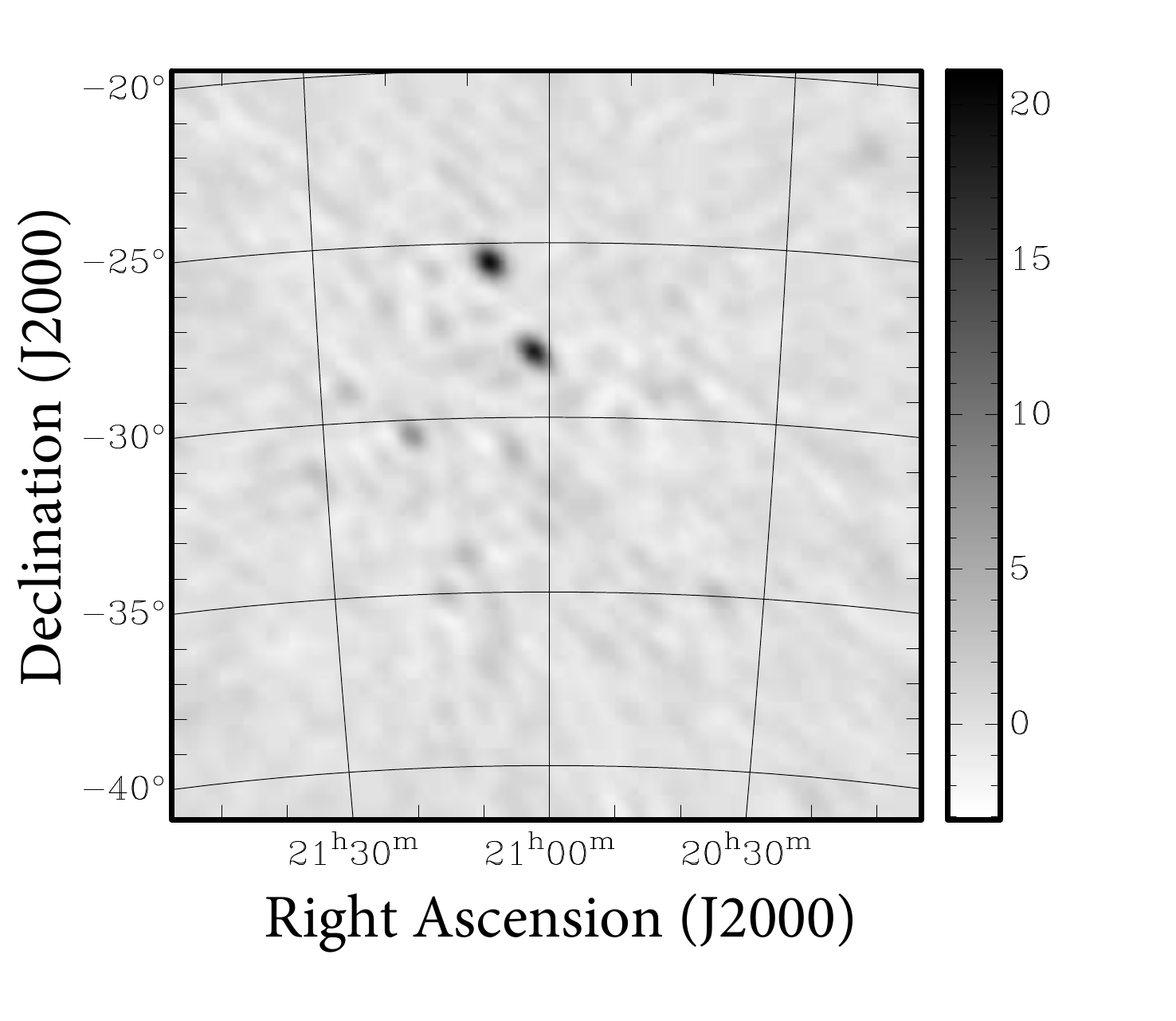}
\includegraphics[width=55mm,angle=0]{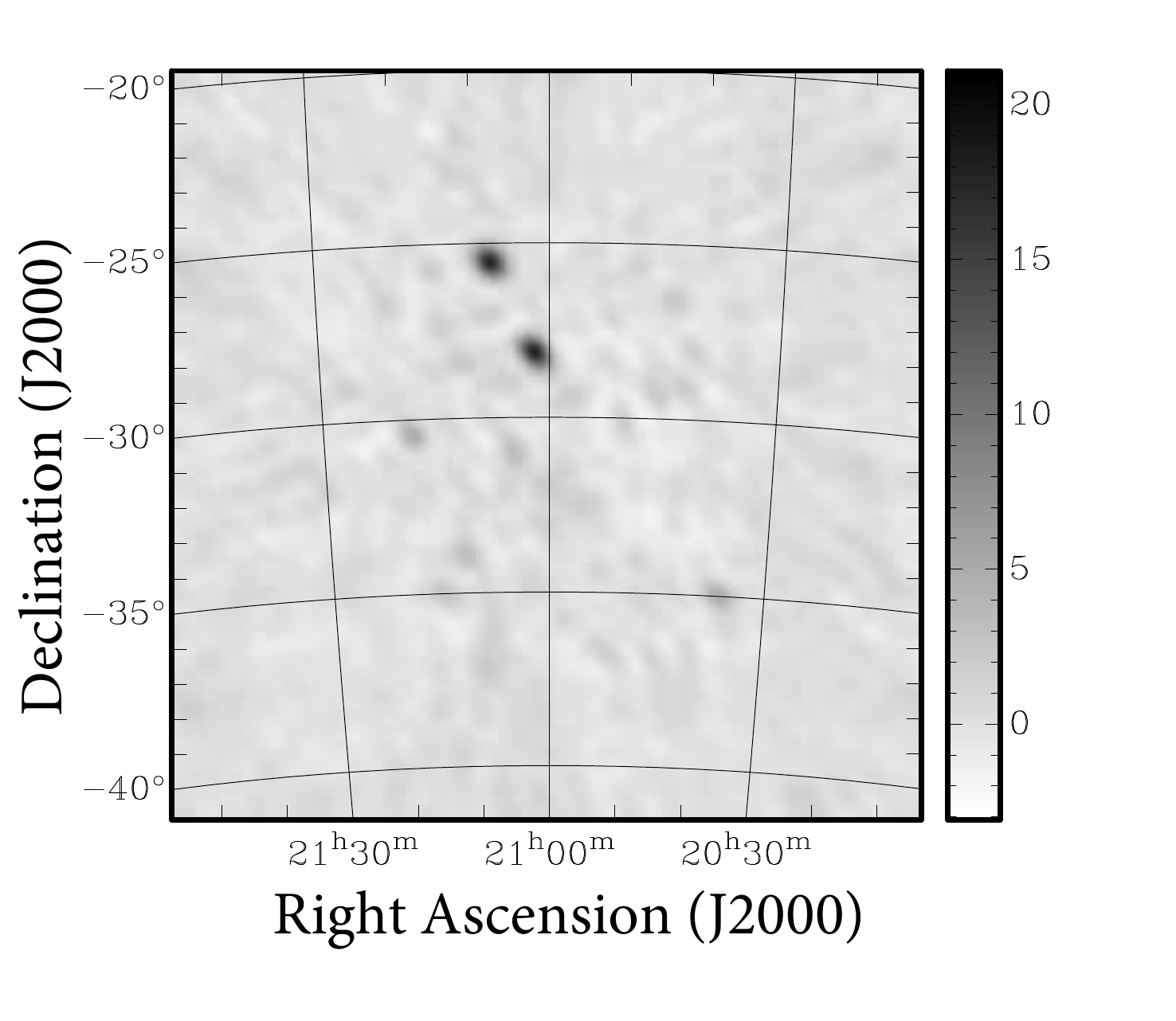}
\includegraphics[width=55mm,angle=0]{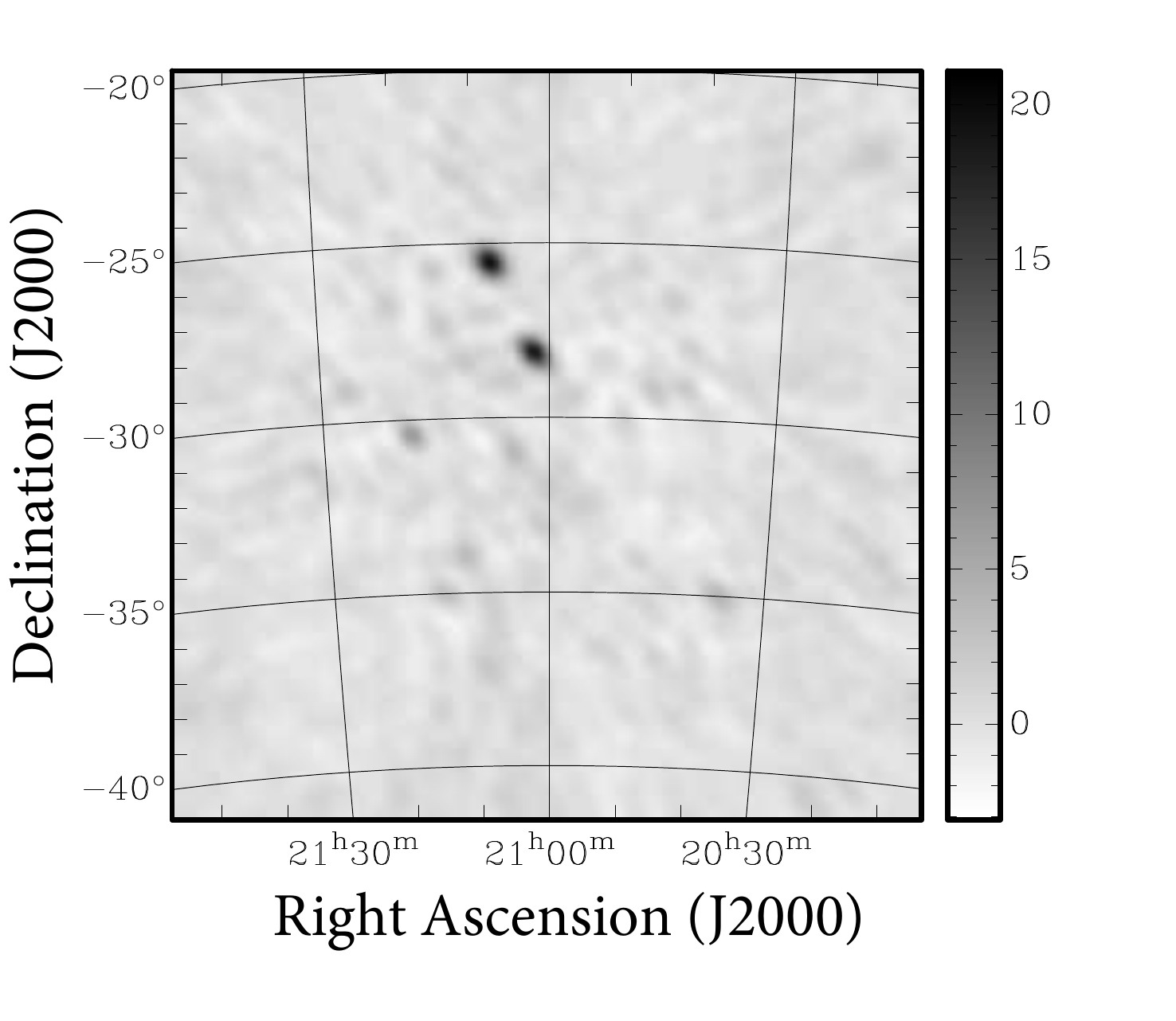}
\includegraphics[width=55mm,angle=0]{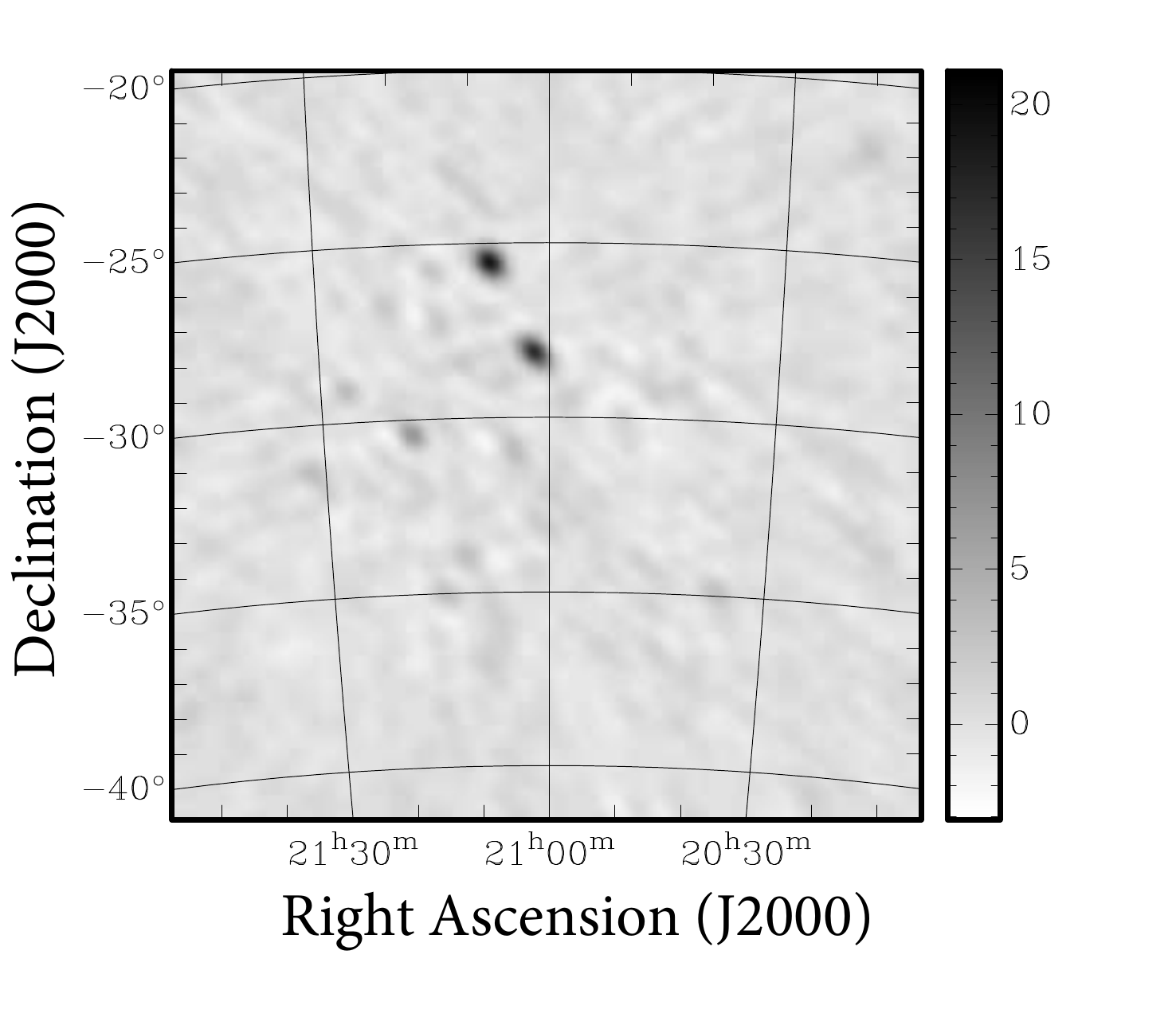}
\includegraphics[width=55mm,angle=0]{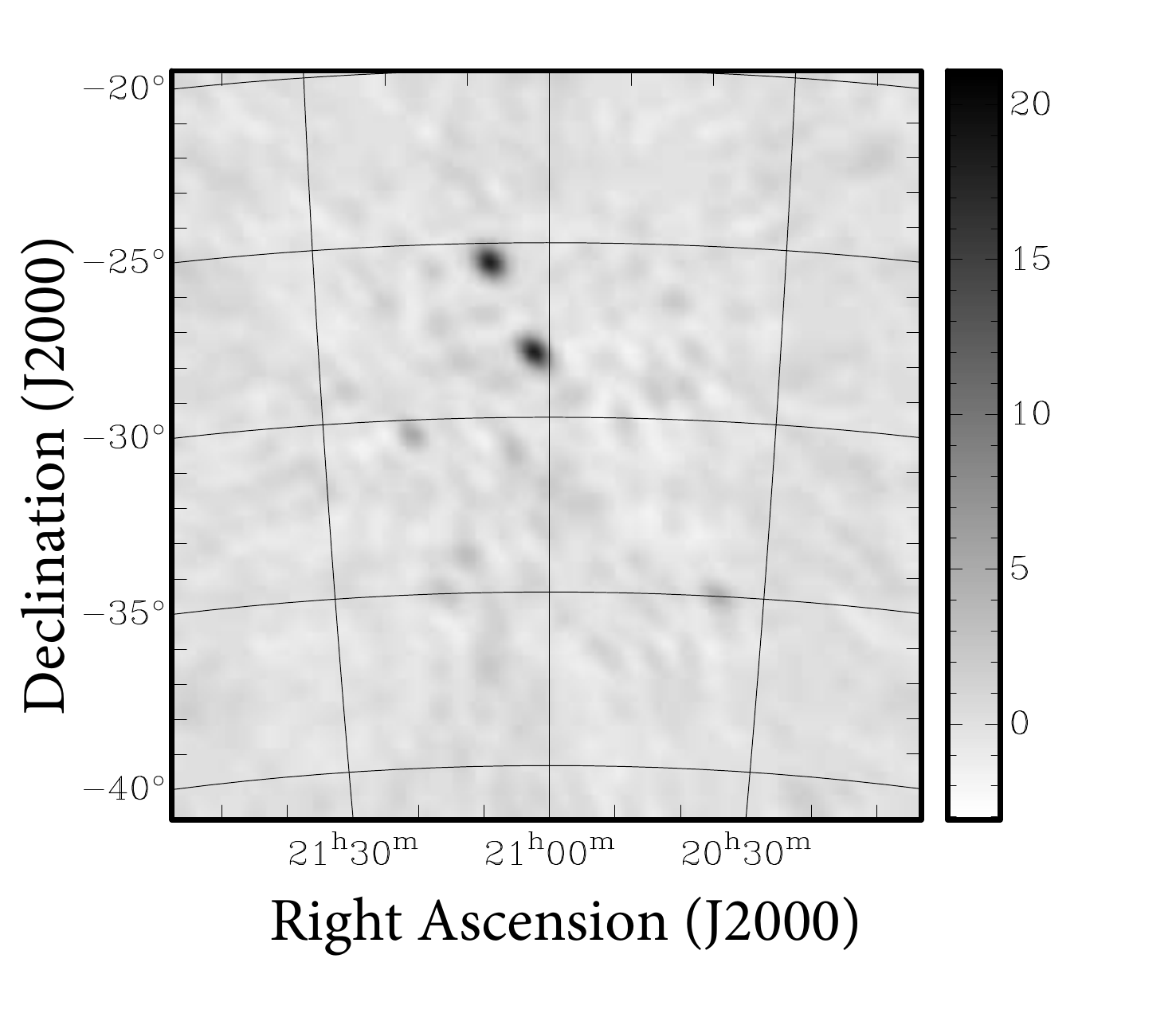}
\includegraphics[width=55mm,angle=0]{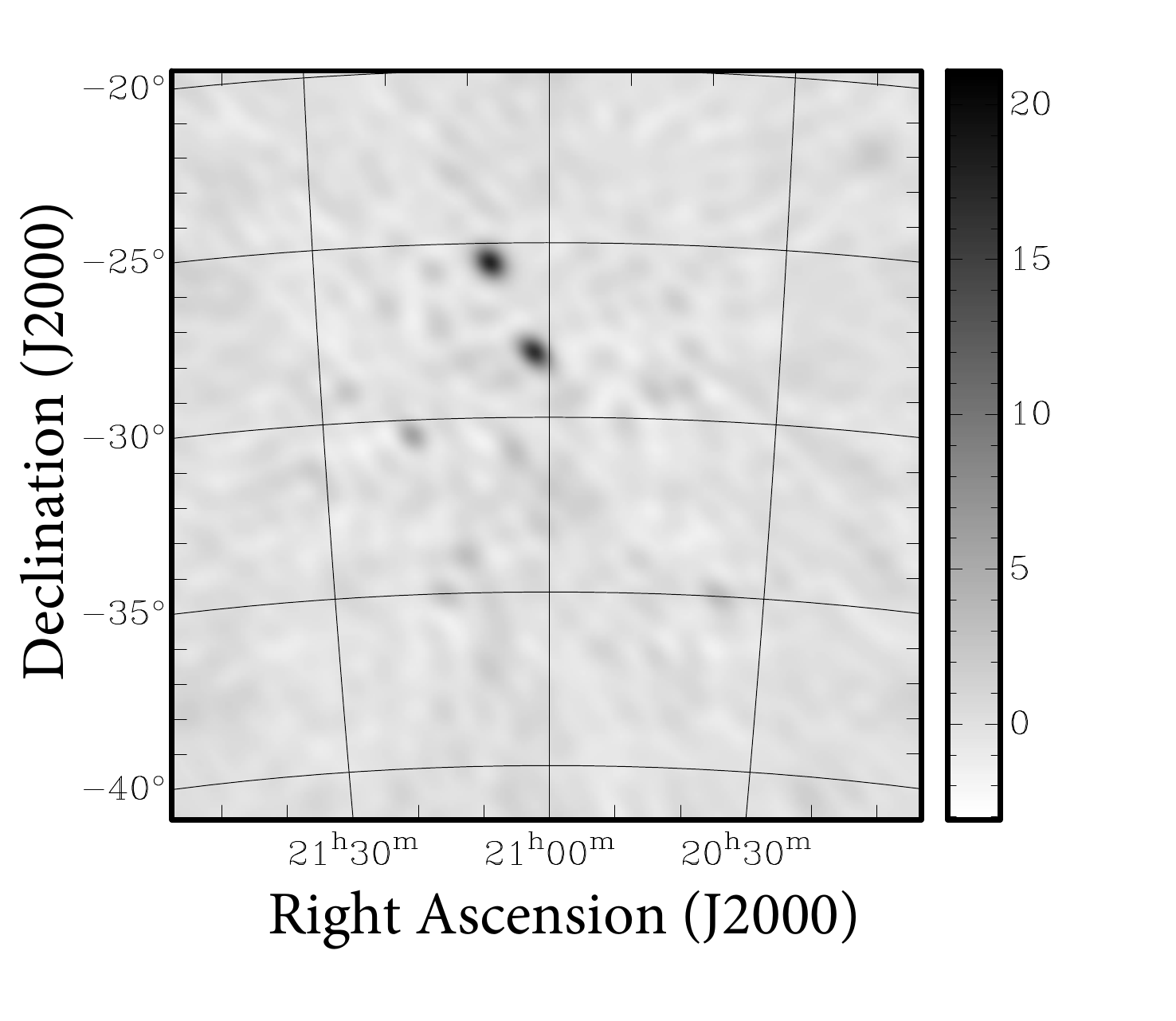}
\includegraphics[width=55mm,angle=0]{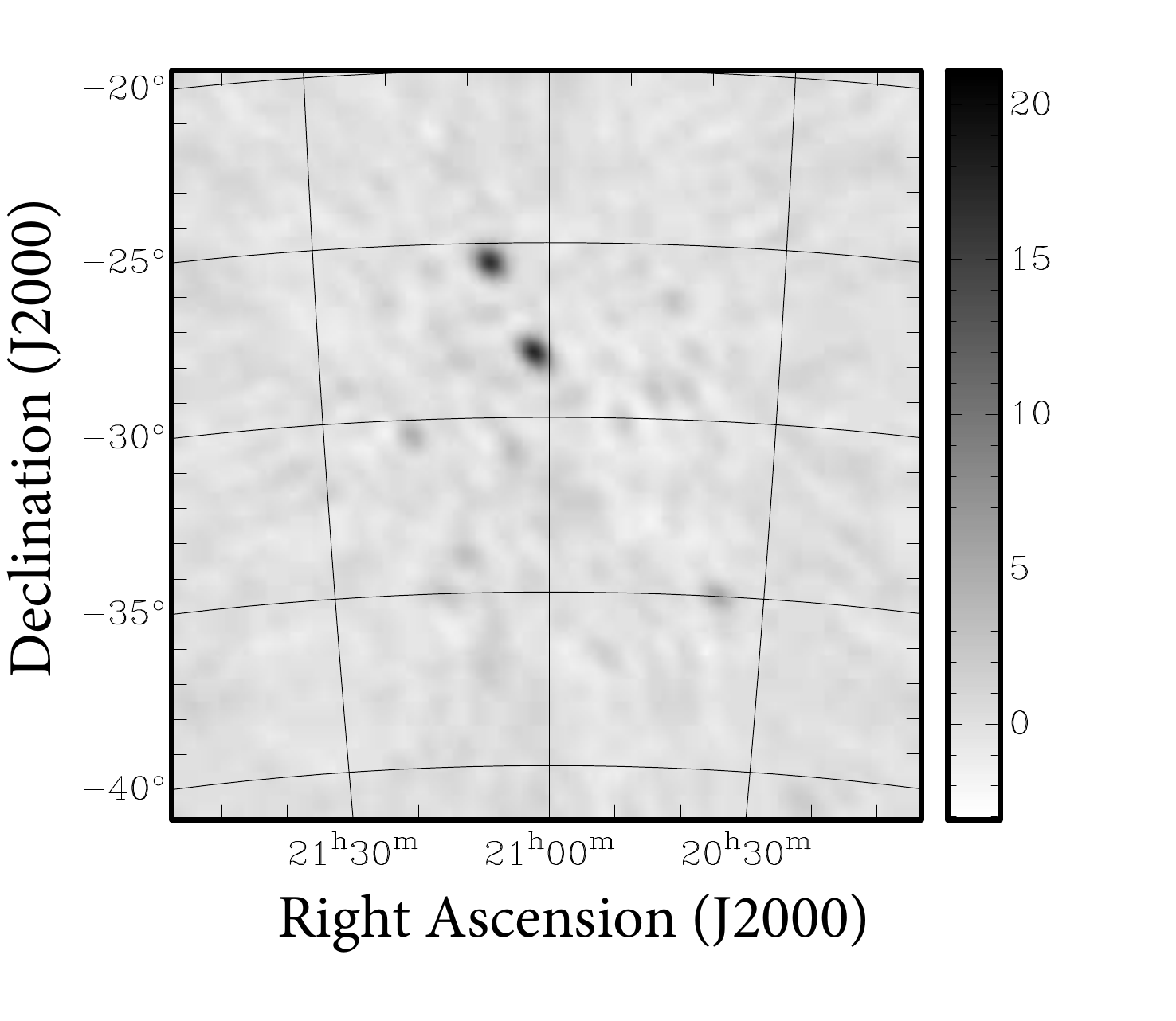}
\includegraphics[width=55mm,angle=0]{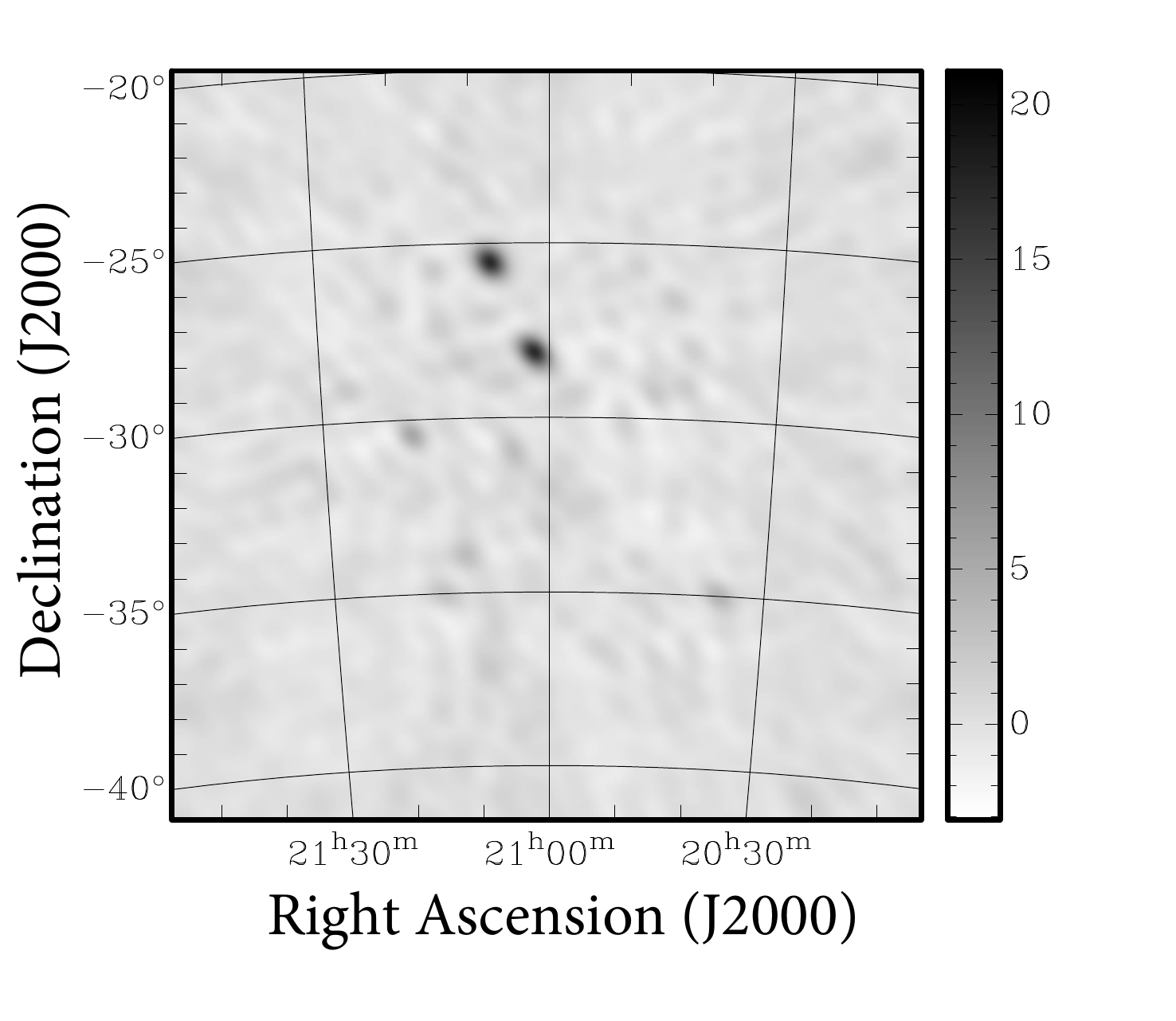}
\caption{This figure shows 10 snapshot images from HERA-47 at 150~MHz. The flux density scale is in Jy~beam$^{-1}$ units.}
\label{fig:10dayimg}
\end{figure*}

\begin{figure*}
\includegraphics[width=55mm,angle=0]{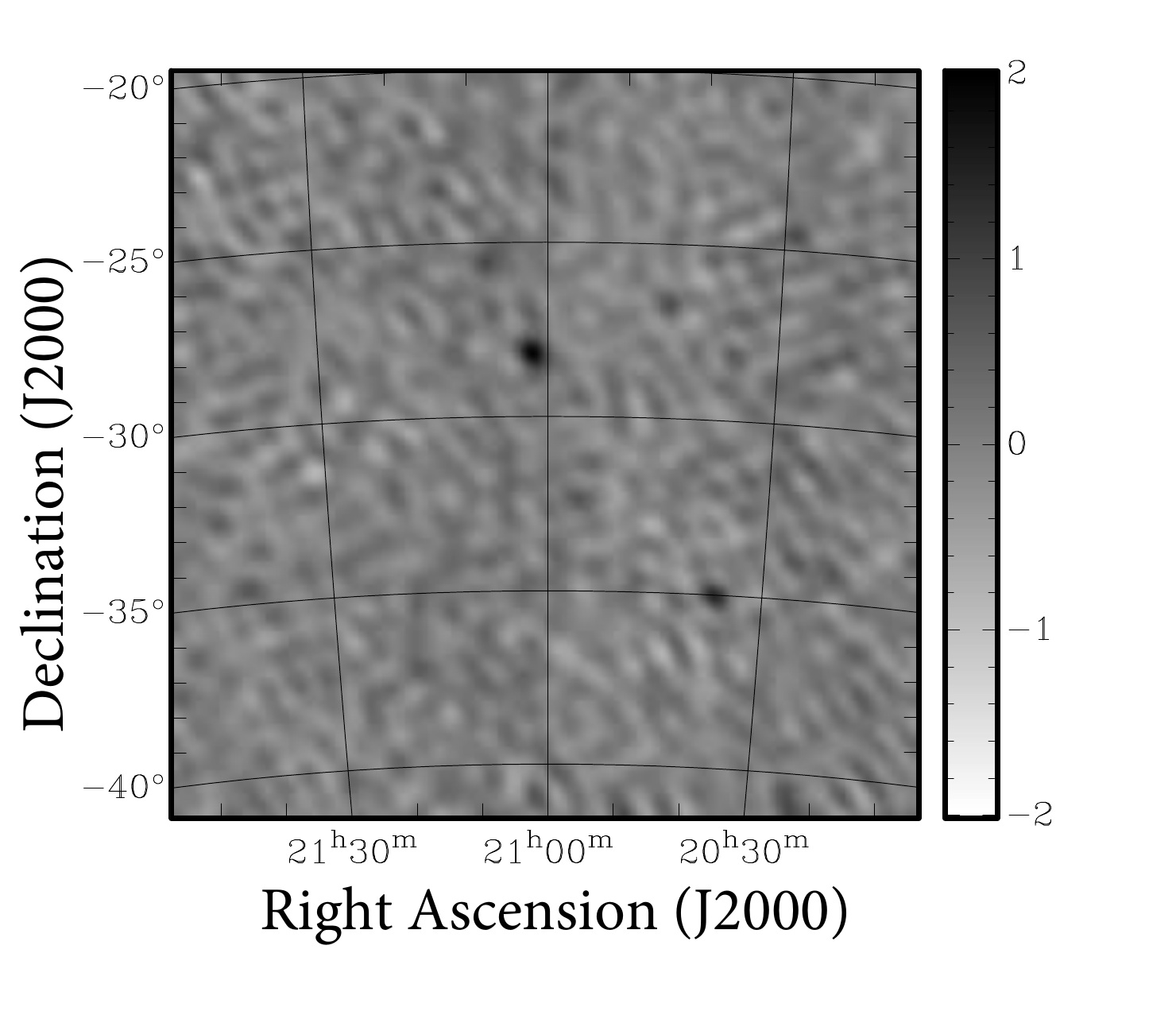}
\includegraphics[width=55mm,angle=0]{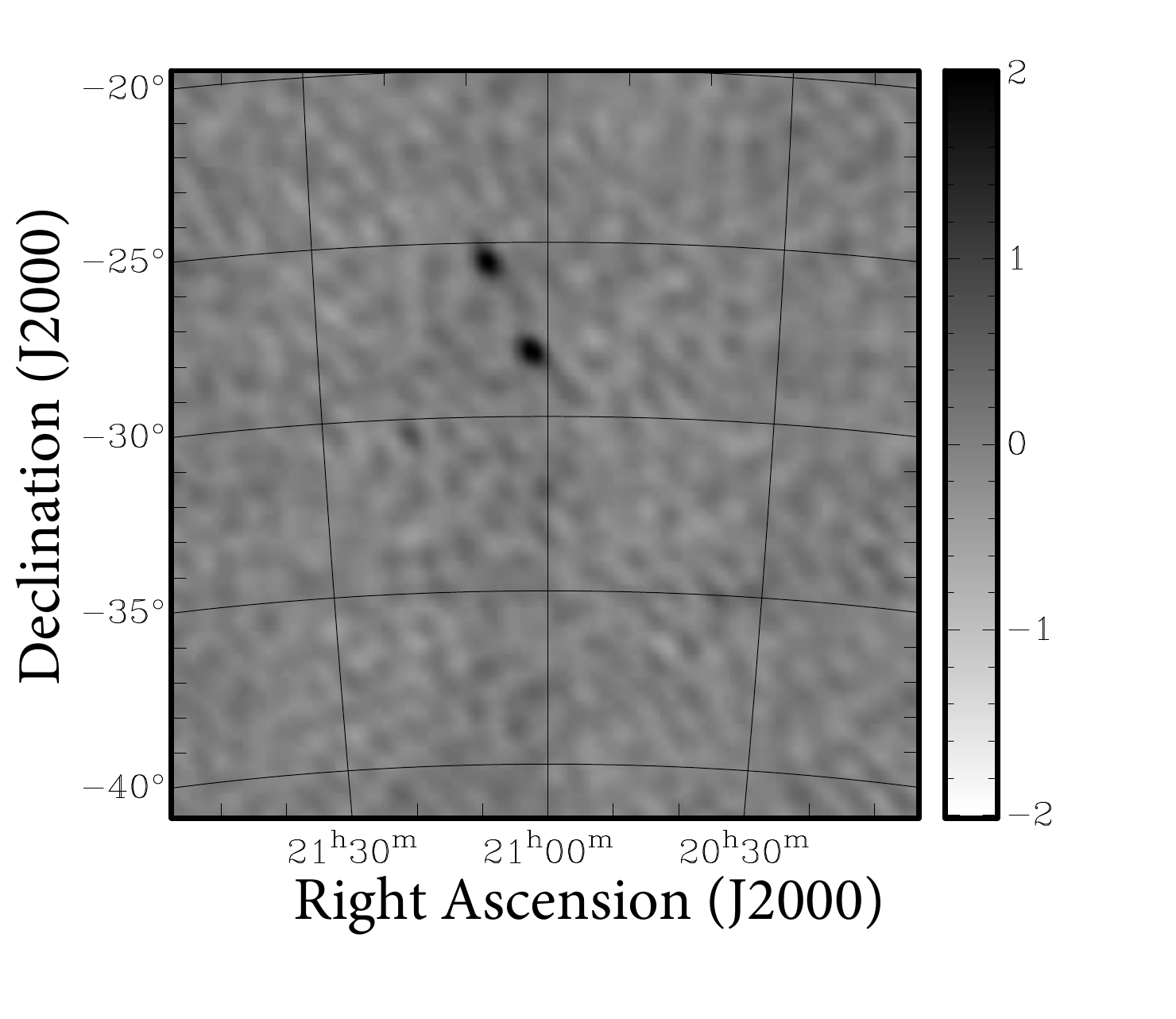}
\includegraphics[width=55mm,angle=0]{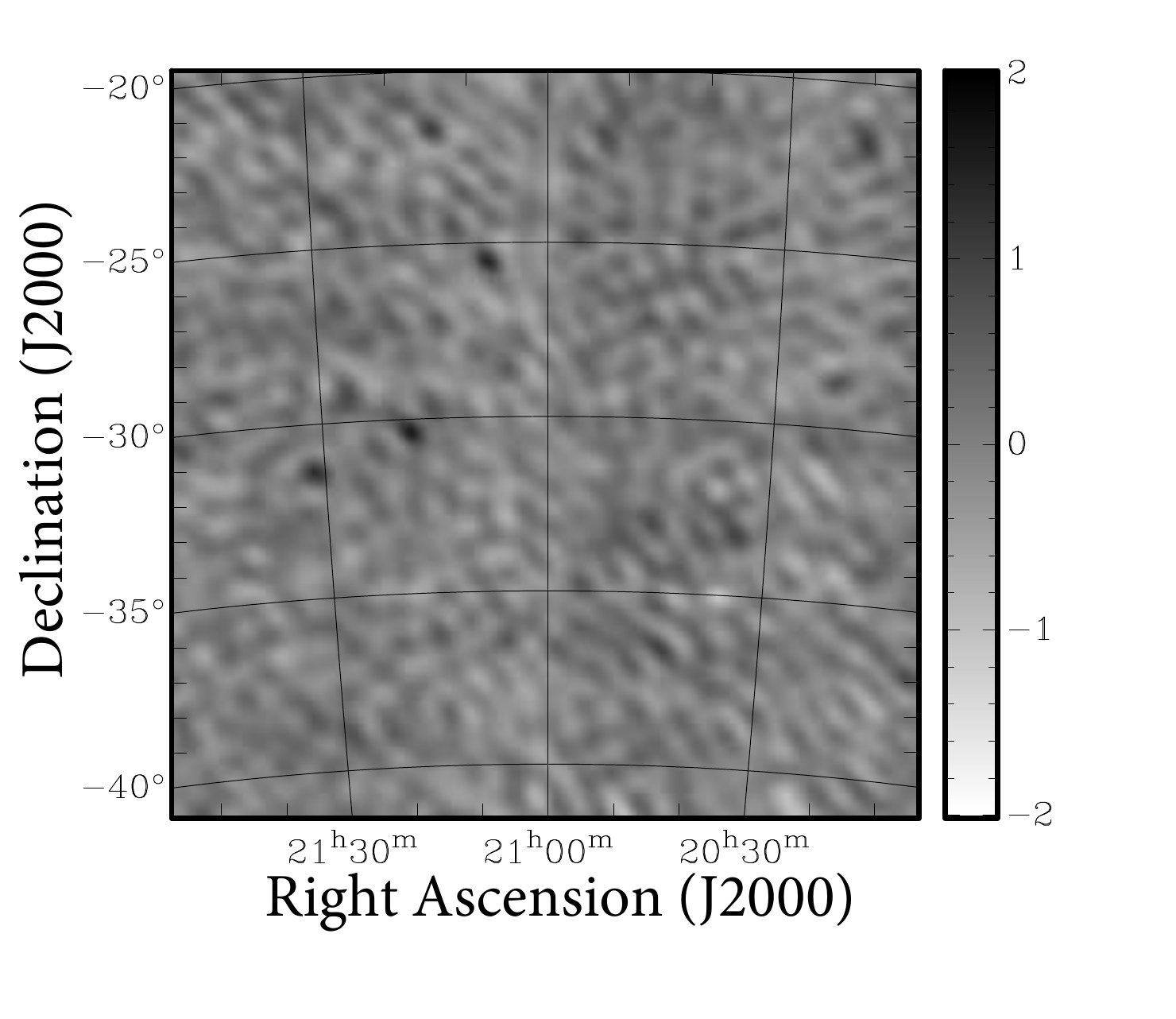}
\includegraphics[width=55mm,angle=0]{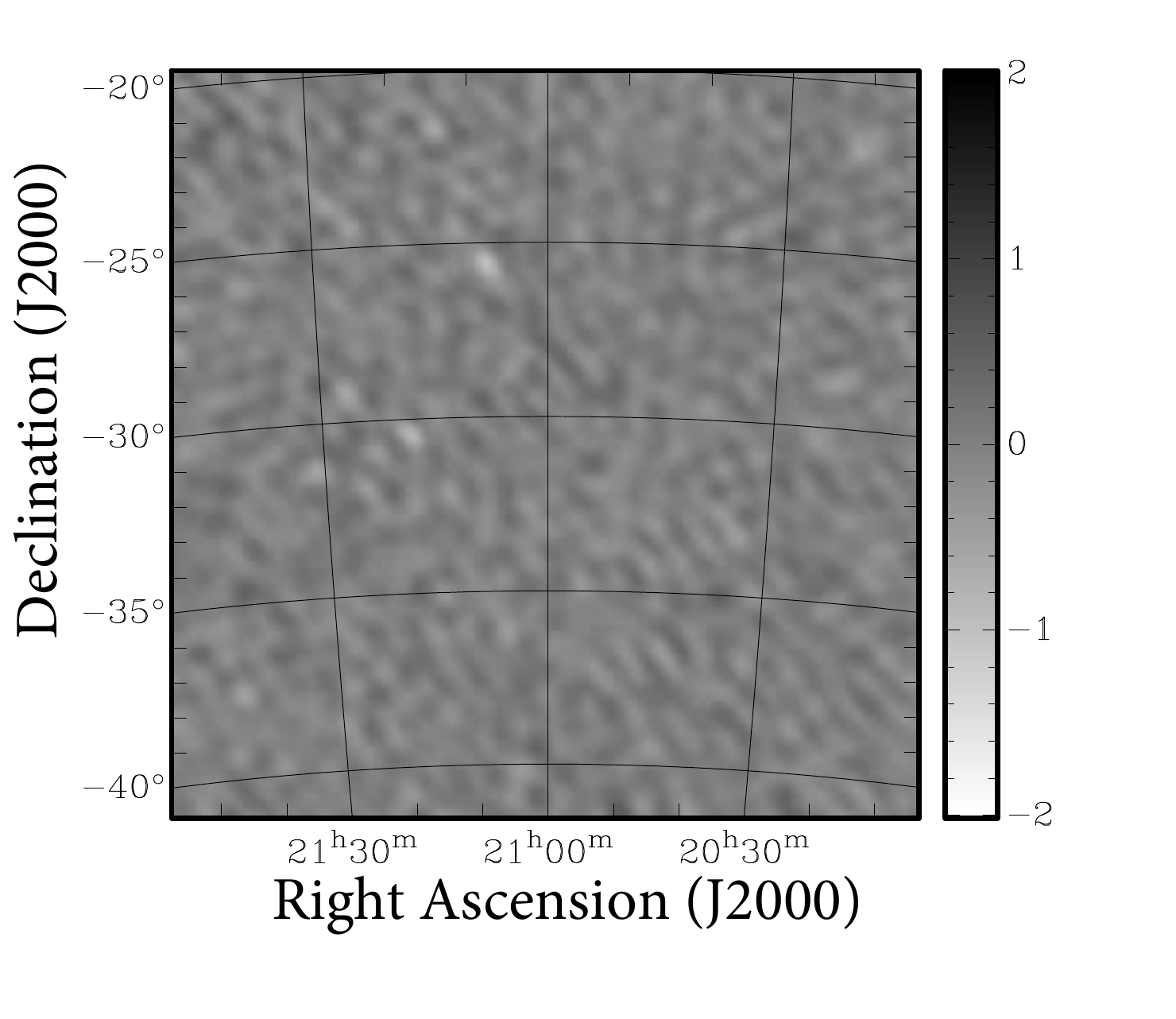}
\includegraphics[width=55mm,angle=0]{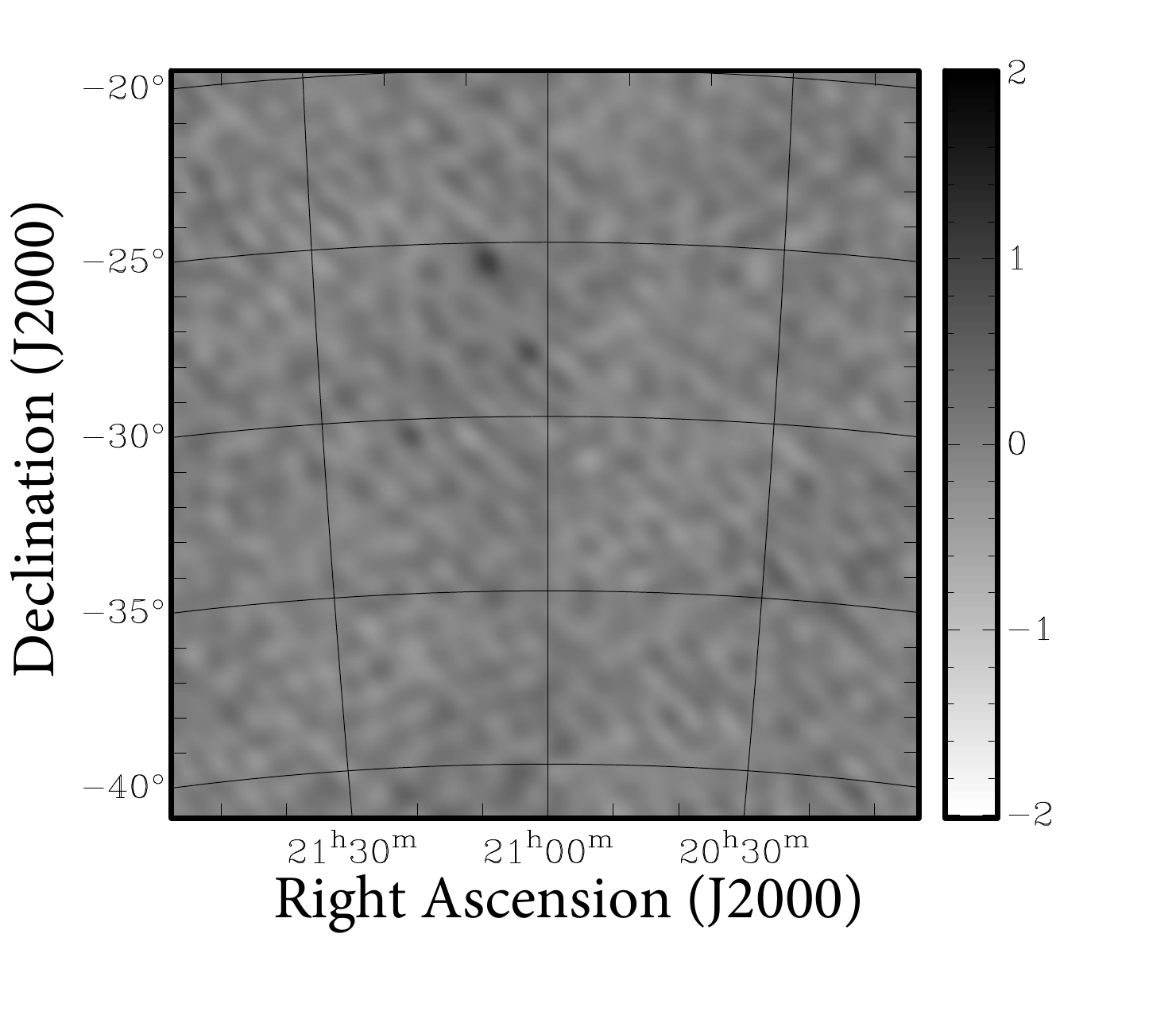}
\includegraphics[width=55mm,angle=0]{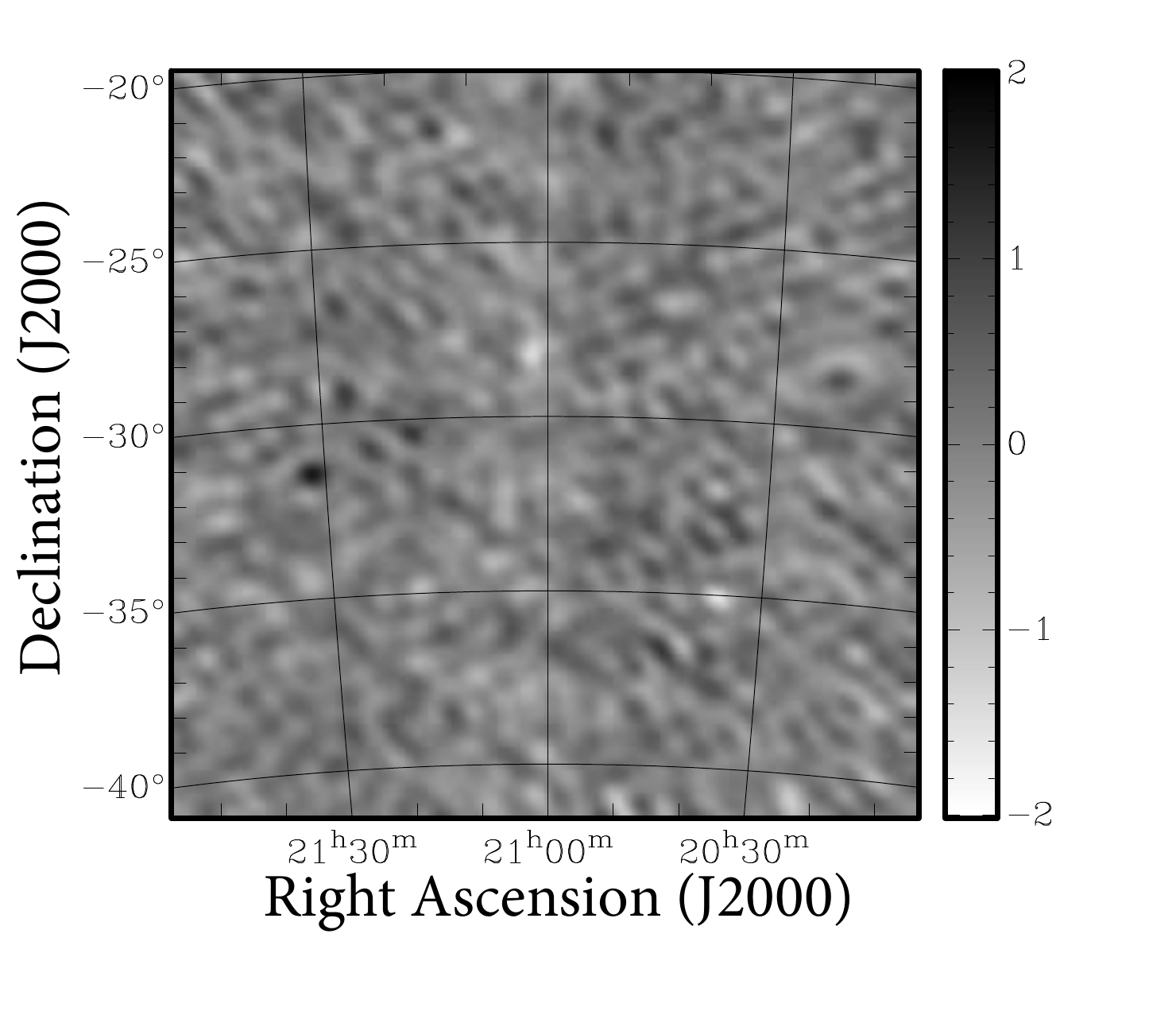}
\includegraphics[width=55mm,angle=0]{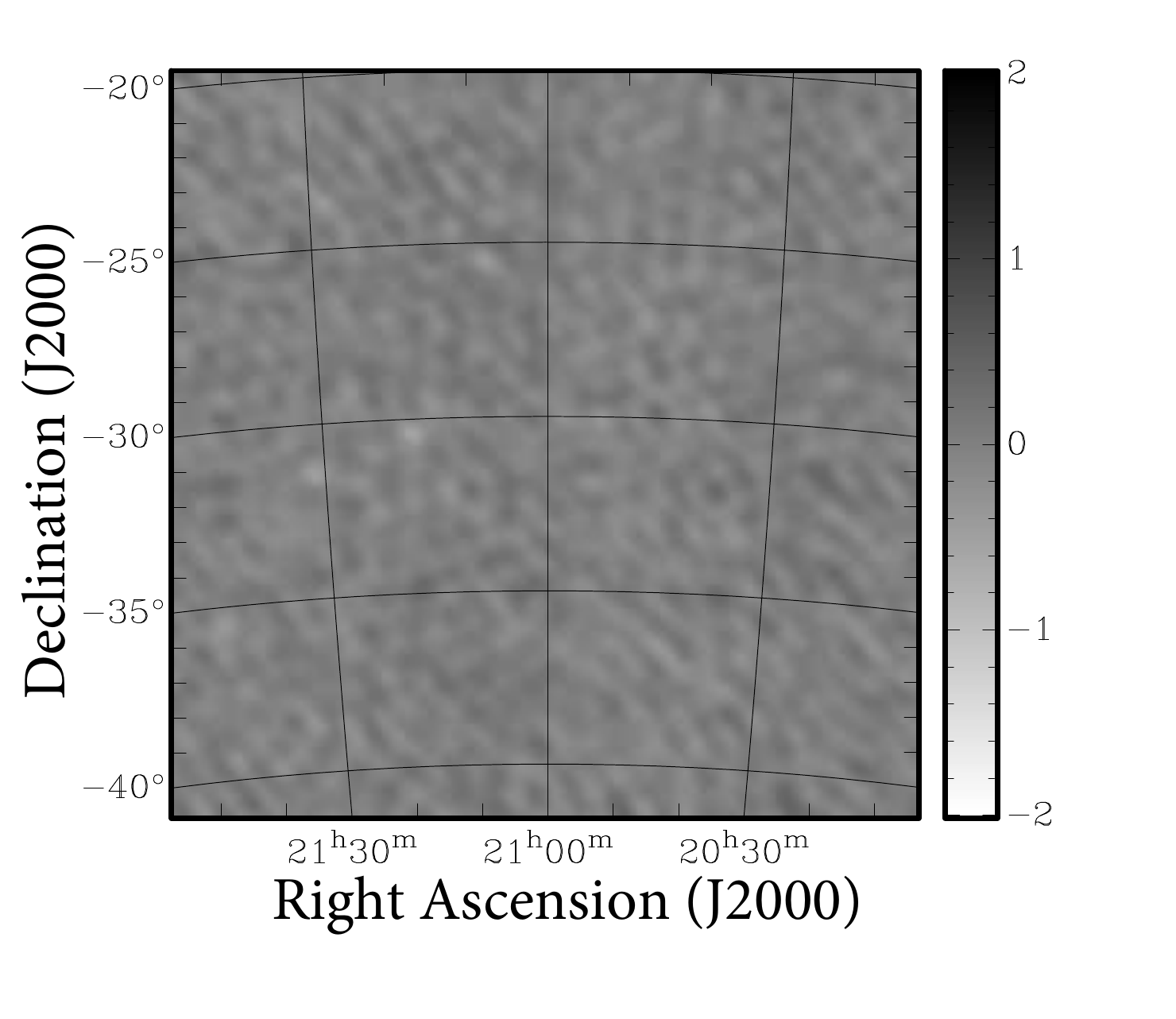}
\includegraphics[width=55mm,angle=0]{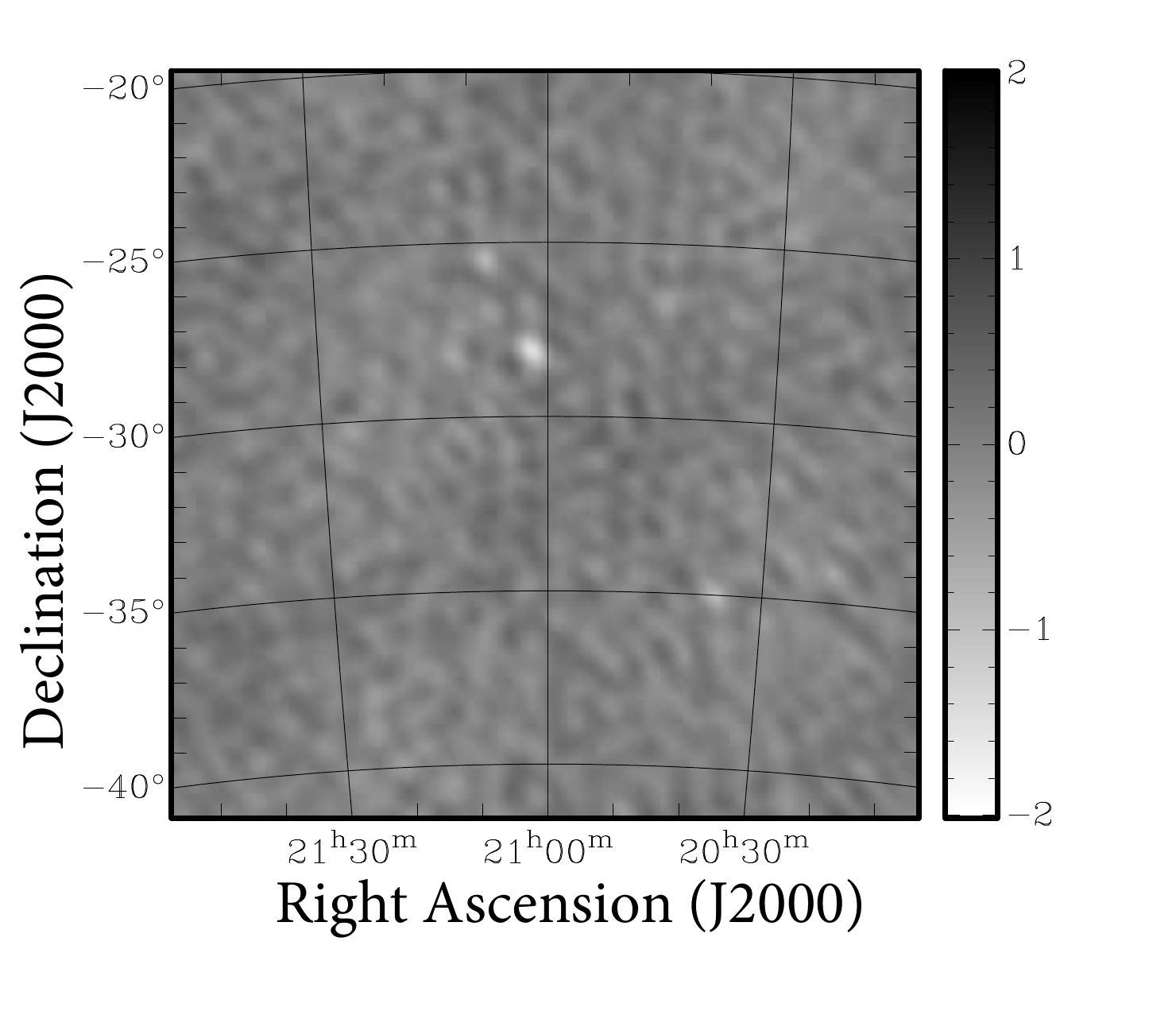}
\includegraphics[width=55mm,angle=0]{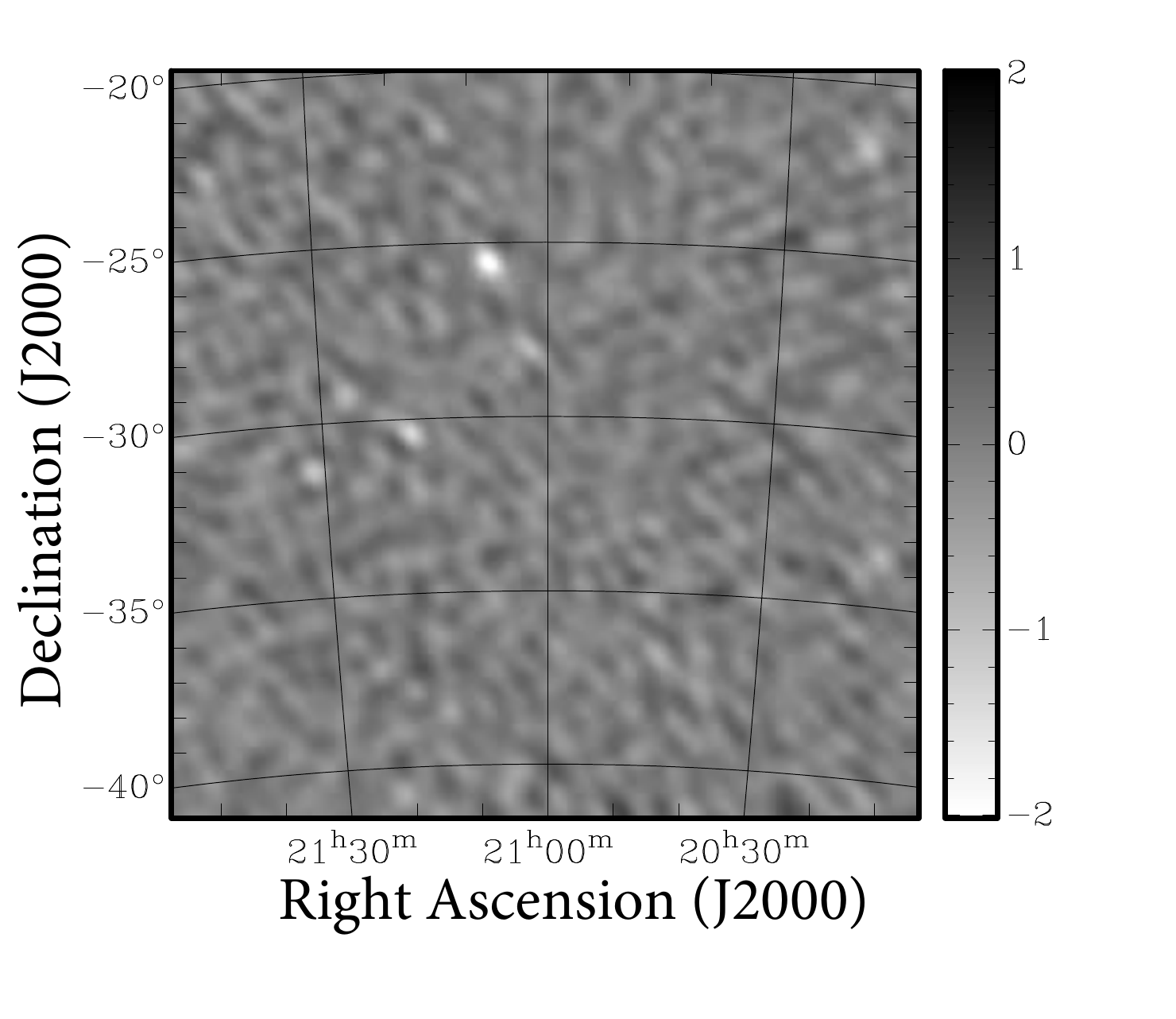}
\includegraphics[width=55mm,angle=0]{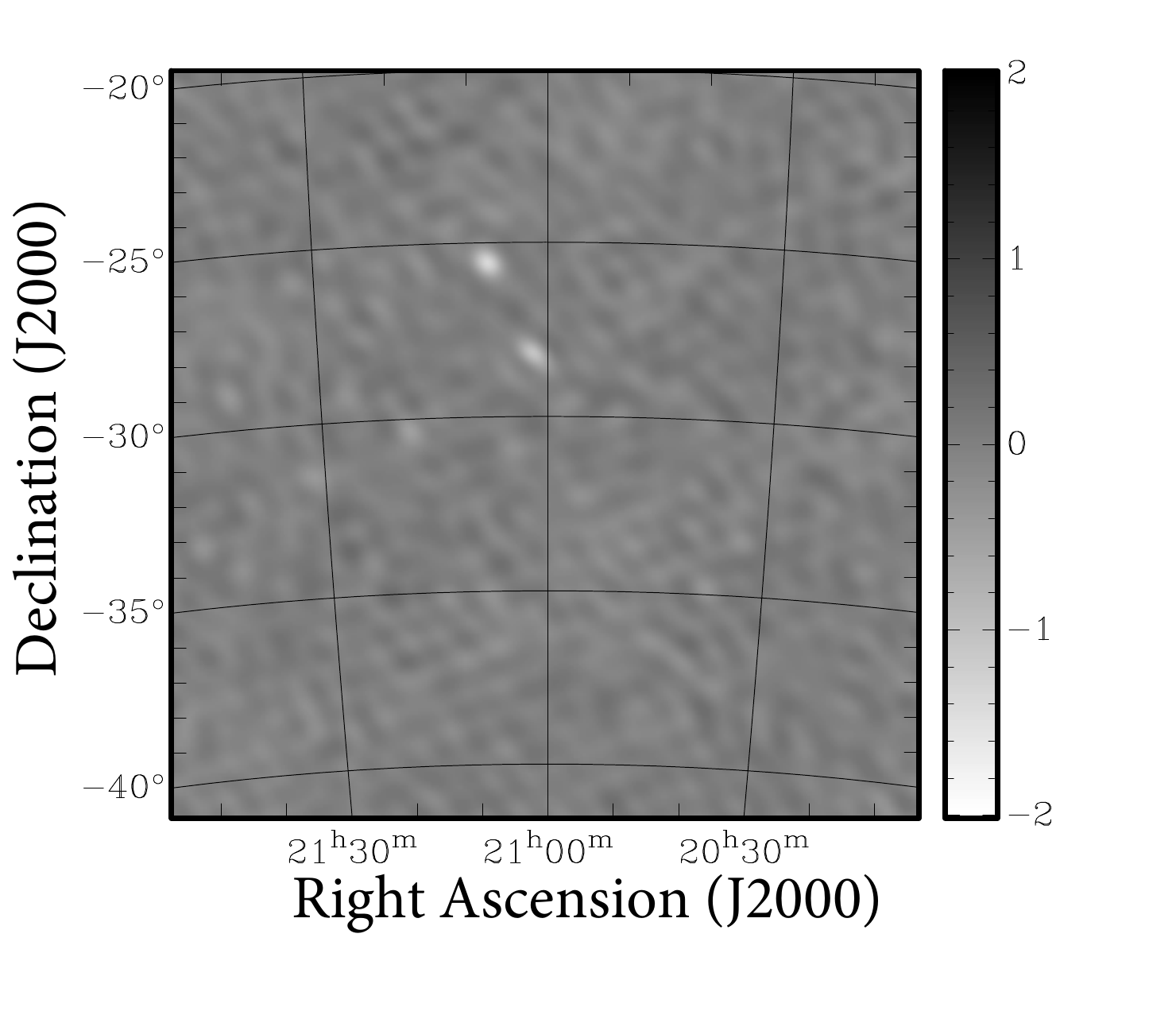}
\caption{This figure display the `difference image' where the mean image has been subtracted out. The flux density scale is in Jy~beam$^{-1}$ units.}
\label{fig:10dayimg_diff}
\end{figure*}

Images of the 10~snapshots are shown in Figure~\ref{fig:10dayimg}. Images at different days are very similar, qualitatively showing good instrumental stability. Image to image variation of the RMS noise in regions of the sky which are mostly empty (away from phase center and void of sources) is between 0.35 and 0.45~Jy~beam$^{-1}$. In these parts of the sky, the primary beam response for the individual fields is considerably lower than the field center and we expect them to be noise dominated. 
As the primary beam response slightly changes based on the transit time at the HERA location, we find that the peak flux density of the images varies up to $5 - 10\%$ over the 10 days (Figure~\ref{fig:10dayimg_diff}), likely due to time variations of the bandpass and imperfect primary beam corrections across snapshots - the primary beam was computed for the first snapshot but observations took place at slightly different LSTs. This variation essentially sets the accuracy of our absolute flux density calibration. We also note that Cygnus~A is above the horizon at the time when observations were taken. Although $\sim 70^\circ$ away from the pointing direction and, therefore, heavily attenuated by the primary beam, it still appears as a source with $\sim 7$~Jy~beam$^{-1}$ peak flux density, possibly affecting the bandpass calibration. We also leave for future work the application of techniques that leverage on the array redundancy to improve calibration \citep{Marthi14,Zhang18,Grobler18,Dillon18,Dillon20}.

%\begin{figure*}
%\includegraphics[width=40mm,angle=0]{figures/042_HHCPF_n.pdf}
%\includegraphics[width=40mm,angle=0]{figures/043_HHCPF_n.pdf}
%\includegraphics[width=40mm,angle=0]{figures/M042_HHCPF_n.pdf}
%\includegraphics[width=40mm,angle=0]{figures/M043_HHCPF_n.pdf}
%\includegraphics[width=40mm,angle=0]{figures/044_HHCPF_n.pdf}
%\includegraphics[width=40mm,angle=0]{figures/045_HHCPF_n.pdf}
%\includegraphics[width=40mm,angle=0]{figures/M044_HHCPF_n.pdf}
%\includegraphics[width=40mm,angle=0]{figures/M045_HHCPF_n.pdf}
%\includegraphics[width=40mm,angle=0]{figures/046_HHCPF_n.pdf}
%\includegraphics[width=40mm,angle=0]{figures/047_HHCPF_n.pdf}
%\includegraphics[width=40mm,angle=0]{figures/M046_HHCPF_n.pdf}
%\includegraphics[width=40mm,angle=0]{figures/M047_HHCPF_n.pdf}
%\includegraphics[width=40mm,angle=0]{figures/048_HHCPF_n.pdf}
%\includegraphics[width=40mm,angle=0]{figures/049_HHCPF_n.pdf}
%\includegraphics[width=40mm,angle=0]{figures/M048_HHCPF_n.pdf}
%\includegraphics[width=40mm,angle=0]{figures/M049_HHCPF_n.pdf}
%\includegraphics[width=40mm,angle=0]{figures/050_HHCPF_n.pdf}
%\includegraphics[width=40mm,angle=0]{figures/051_HHCPF_n.pdf}
%\includegraphics[width=40mm,angle=0]{figures/M050_HHCPF_n.pdf}
%\includegraphics[width=40mm,angle=0]{figures/M051_HHCPF_n.pdf}
%\caption{The two left columns show 10 snapshot images from HERA-47 at 150~MHz. The two right columns display the `difference image' where the mean image has been subtracted out. The flux density scale is in Jy~beam$^{-1}$ units.}
%\label{fig:10dayimg}
%\end{figure*}

\subsection{LST binning \& SEFD evaluation}
\label{sec:lstbin_sefd}
We bin each night of visibility data in LST. We chose a 2~minute bin resolution, such that we can minimize the variation of the primary beam. For each observing night and LST bin, we only average redundant baselines (e.g. baselines of the same length and orientation). 
This ensures that we are coherently averaging the baselines and not mixing up emissions from the sky as the earth rotates. 

\begin{figure}
\includegraphics[scale=0.6]{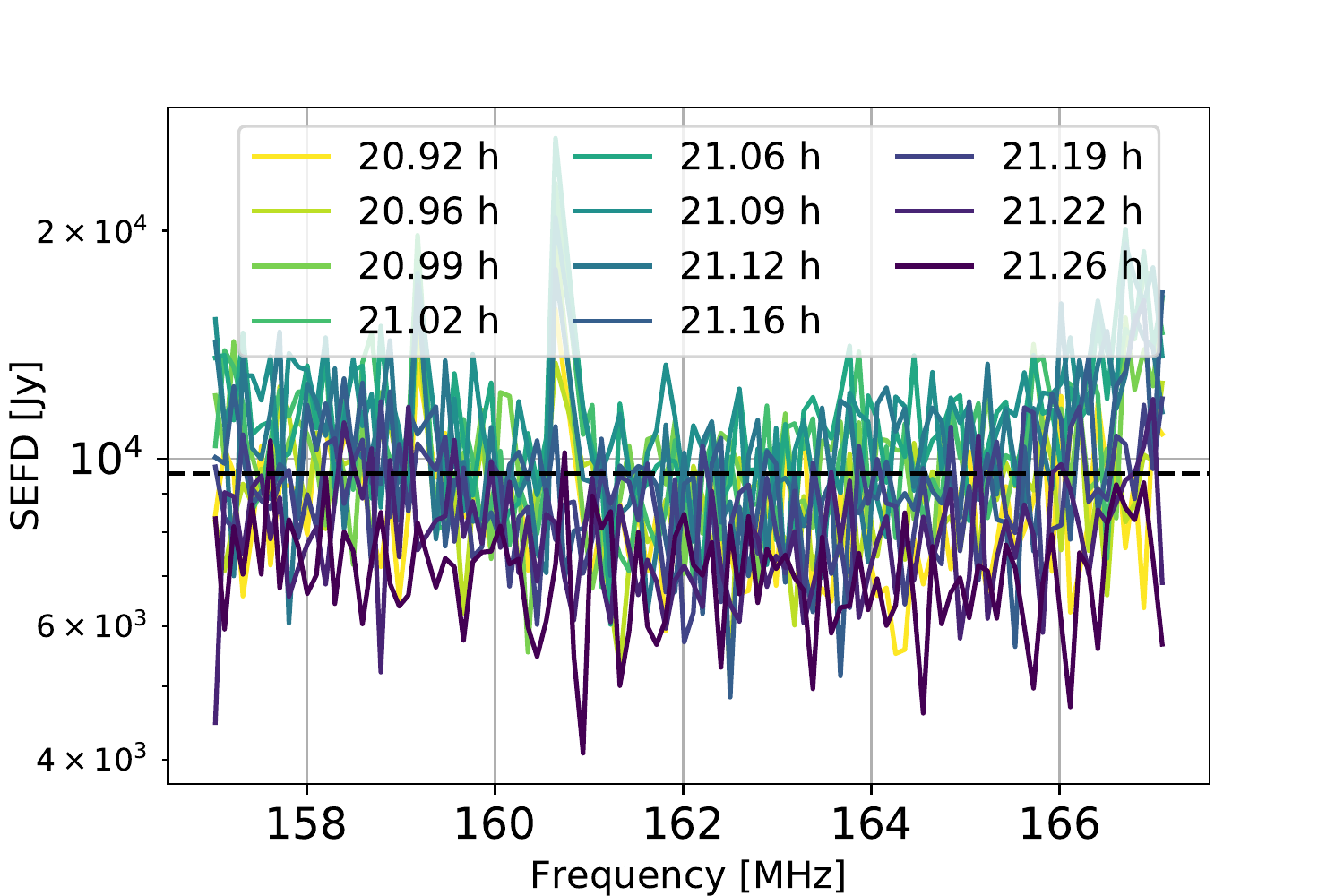}
\caption{Estimated SEFD as a function of frequency for the different LST bins. The dashed line shows the mean value for all the LST bins and frequency channels.}
\label{fig:sefd_freq}
\end{figure}

We empirically estimate the System Equivalent Flux Density (SEFD) of the different LST combined visibility data sets by taking the difference of two adjacent frequency channels \citep{Patil16}. This difference can be used to estimate the noise RMS, $\sigma(u, v, \nu)$. For each polarisation, we have \citep{Thompson17}:
\begin{equation}
\sigma(u, v, \nu) = \frac{1}{\sqrt{N_{\rm vis}(u, v, \nu)}} \frac{\rm{SEFD(\nu)}}{
\sqrt{\Delta \nu\, \Delta t}},
\label{eq:sefd}
\end{equation}
where $\Delta \nu$ and $\Delta t$ are the frequency channel width and integration time respectively. 
We use equation~\ref{eq:sefd} to estimate the SEFD for the different LST bins as a function of frequency (Figure~\ref{fig:sefd_freq}). The factor $N_{\rm vis}(u, v, \nu)$ in equation~\ref{eq:sefd} is the number of redundant visibilities. We find a SEFD~$\sim (9.5 \pm 2.4) \times 10^3$~Jy (here, the mean and the uncertainty is estimated from all the LST bins and frequency channels in Figure \ref{fig:sefd_freq}) for the $157.03 - 167.09$~MHz range which we use for the power spectrum analysis. In temperature units, this is equivalent to $\sim 327 \pm 84$~K around a central frequency of 162~MHz. We use a scaling factor of $(10^{-26} \, \lambda^2) / (2 \, {\rm k_{B}} \, \Omega_{{\rm P}})$, where $\Omega_{{\rm P}}$ is the angular area of the primary beam \citep{Parsons_normmemo17}, to convert from Jy to K. The estimated SEFD values are consistent with the HERA system temperature derived by using differences of visibility spectra for sky-calibrated data for a fixed LST on two consecutive days \citep{Carilli17}.

Visibilities observed at the same LST time should ``see" the same sky. Assuming that over the 2~min LST bin the change in the primary beam is not significant, all the 2-min averaged visibilities corresponding to similar LST bins are therefore also coherently averaged (after baselines of same length and slope have been averaged). Visibility data sets from different LST bins, on the other hand, correspond to different parts of the sky and therefore cannot be coherently averaged. However, the 21-cm signal power spectra should only depend on baseline length, not time. We can thus incoherently combine them when producing power-spectra (e.g. we average the power spectrum from different LST bins).
In the following subsection, we describe our power spectrum estimation procedure. We focus our discussion on the line of sight and delay power spectrum in the $k_{\perp} - k_{\parallel}$ plane or, equivalently, baseline - delay plane.

\subsection{Delay power spectrum}
\label{sec:estps}
Intrinsic flat spectrum sky emission appears as a Dirac delta function in delay space, where the Fourier transform along the frequency axis (delay transform) acts as a one-dimensional, per-baseline ``image" \citep{parsons12a}.
Smooth-spectrum foregrounds are bound by the maximum geometric delay that depends upon the baseline length. We investigate such foreground isolation via the `delay spectrum' $\hat{V}(\mathbf{u}, \tau)$, defined as the inverse Fourier transform of $V(\mathbf{u}, f)$ along the frequency coordinate \citep{parsons12a, parsons12b}: 
\begin{equation}
\label{eqn:delay-transform}
\hat{V}(\mathbf{u}, \tau) \equiv \int V(\mathbf{u}, f)\,W(f)\,e^{i2\pi f\tau}\,\dif f
\end{equation}
where, $W(f)$ is a spectral window function \citep{Vedantham12, Thyagarajan13, Choudhuri16} and $\tau$ represents the signal delay between antenna pairs $\tau = \frac{\mathbf{u}\cdot\hat{\boldsymbol{s}}}{c}$, where, $\mathbf{u}$ is the baseline vector towards the direction $\hat{\boldsymbol{s}}$ and $c$ is the speed of light. We finally
squared the visibilities, $\hat{V}(\mathbf{u}, \tau)$, to form the delay power spectrum. Unlike an image based estimator where the upper and lower frequencies incorporate information from baselines of different physical length, the delay power spectrum respects baseline migration, i.e., the same baselines contribute to all frequencies \citep[e.g.,][]{Morales12}. In our analysis we used baselines with length $|\mathbf{u}| \le 60 \,{\rm m}$ and a non-uniform discrete Fourier transform to compute the line-of-sight delay transform of the visibilities in order to take proper account of the flagged frequency channels. We choose a Blackman window function which offers a $\sim -67$~dB side lobe suppression. 

For a single baseline, we can estimate the delay power spectrum (e.g. the cylindrical power spectrum) as~\citep{parsons12a}:
\begin{equation}
P(k_{\perp}, k_{\parallel}) \approx \big(\frac{10^{-26} \lambda^2}{2 \, {\rm k_B}} \big)^2 \times \frac{X^2 Y}{\Omega_{\mathrm{PP}} B} \left|
\hat{V}(\mathbf{u}, \tau)\right|^2
\end{equation}
where, $\lambda$ corresponds to the wavelength of the mid-frequency of the band, ${\rm k_B}$ is the Boltzmann constant, $B$ is the bandwidth, $\Omega_{\mathrm{PP}}$ is the angular area of the primary beam and X, Y are conversion factors from angle and frequency to co-moving scales. As discussed, the power spectrum is averaged over all LST bins. Moreover, we also average over all $k$ modes with the same $k_{\perp}$, i.e., the modes which have the same baseline length. We used the power-square beam from the HERA beam measurements\footnote{\url{https://github.com/HERA-Team/hera-cst}}\textsuperscript{,}\footnote{\url{http://reionization.org/science/memos/}}\citep{Parsons_normmemo17} to estimate the beam area \citep[equation~B10 in][]{Parsons14}. The power spectrum has units of ${\rm K^{2} \, [h^{-3}\, cMpc^{3}}]$.
Fourier modes $(k_{\perp},k_{\parallel})$ are in units of inverse co-moving distance and are given
by \citep[e.g.,][]{Morales06,Trott12}:
\begin{eqnarray}
k_{\perp} &=& \frac{2 \pi |\mathbf{u}|}{D_M(z)},\\ 
k_{\parallel} &=& \frac{2 \pi H_0 \nu_{21} E(z)}{c(1+z)^2} \tau,\\
k &=& \sqrt{k_{\perp}^2 + k_{\parallel}^2}
\end{eqnarray}
where $D_M(z)$ is the transverse co-moving distance, $H_0$ is the Hubble constant, $\nu_ {21}$ is the frequency of the hyperfine transition, and $E(z)$ is the dimensionless Hubble parameter~\citep{Hogg99}. 

\begin{figure}
\includegraphics[scale=0.55]{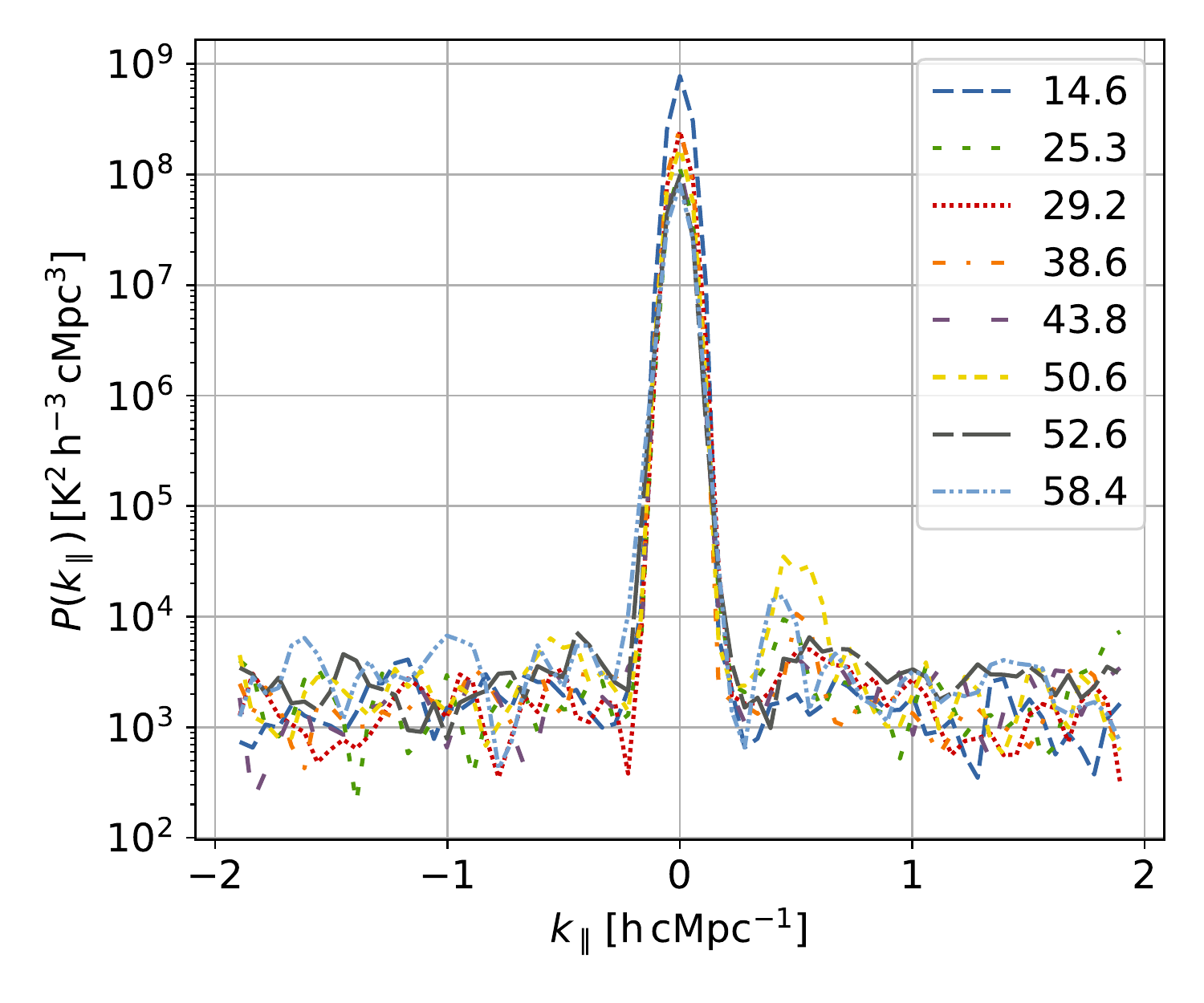}
%\plotone{1dPS_HERA47_BAGPR.pdf}
\caption{Power spectrum of a 2~min LST binned data as a function of $k_{\parallel}$ for the baselines included in our analysis. The baselines in units of meters are shown in the legend.}
\label{fig:1d_I_ps_bagpr}
\end{figure}

Figure~\ref{fig:1d_I_ps_bagpr} shows the delay power spectrum for a 2 min LST binned data (e.g. we consider one LST bin only) as a function of $k_{\parallel}$, up to $\tau \sim 3.6$~$\mu$s, corresponding to $k_{\parallel} \sim 2.0$~h~cMpc$^{-1}$. We used a $10$~MHz bandwidth centred at $162.06$~MHz to estimate the delay spectrum. We found most of the foreground power is confined within $k_{\parallel} \le 0.2$~h~cMpc$^{-1}$ and foreground excess beyond that is largely limited for most baselines, however, there is some signature of a signal with a $\sim 1$~MHz period  \citep{Kern19b}, corresponding to $k_{\parallel} \sim 0.5$~h~cMpc$^{-1}$, for all the baselines considered.

\begin{figure}
\includegraphics[scale=0.5]{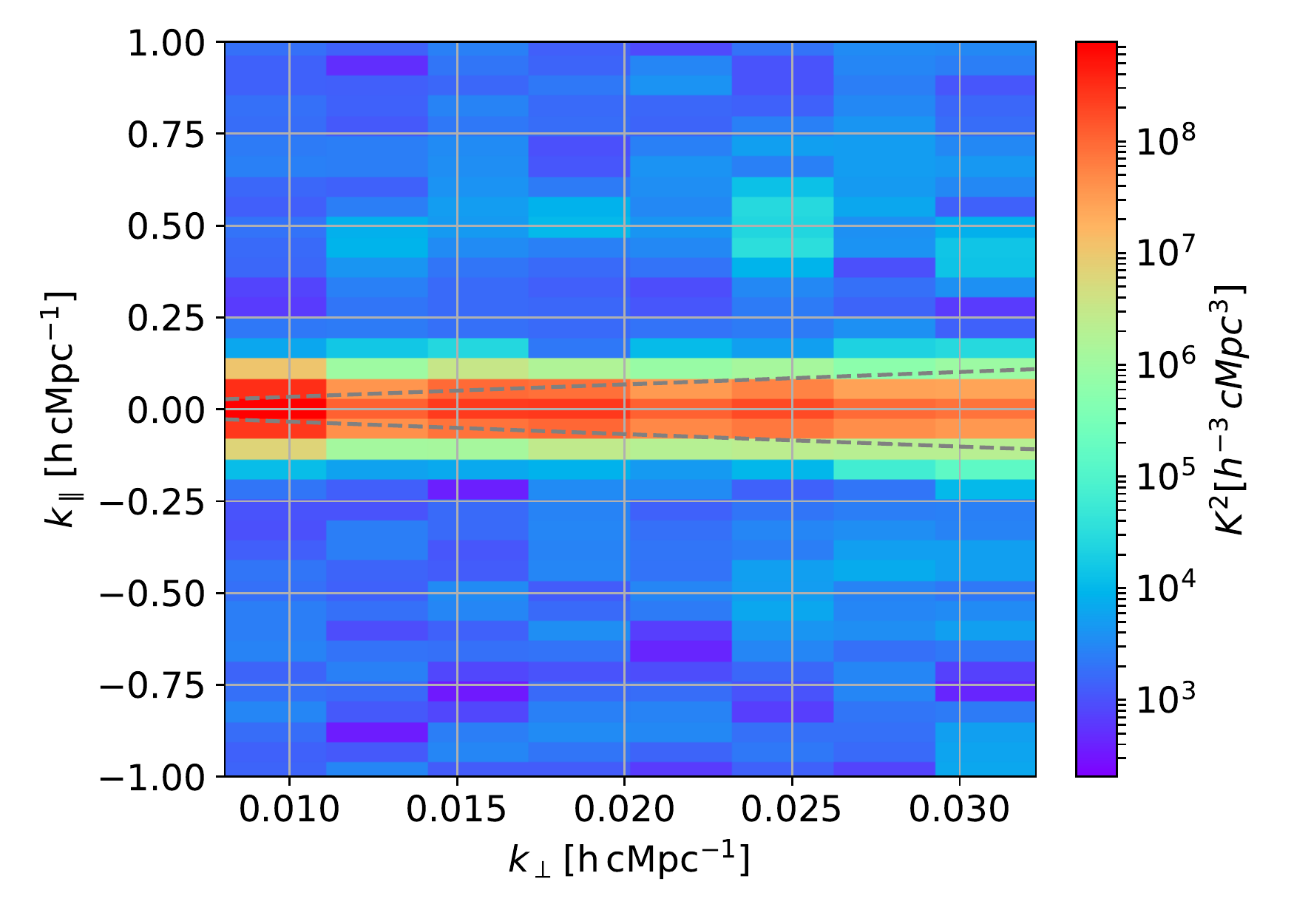}
%\includegraphics[scale=0.5]{2dPS_HERA47_BAGPR_Ratio20percent.pdf}
%\plotone{2dPS_HERA47_BAGPR.pdf}
\caption{Delay power spectrum in the ($k_{\perp} - k_{\parallel}$) plane for a 2~min LST binned data. 
%of JD2458048. 
The dashed line represents the horizon line corresponding to $\theta_{\rm max} = 90^{\circ}$.}
\label{fig:2d_I_ps_bagpr}
\end{figure}

Figure~\ref{fig:2d_I_ps_bagpr} presents the delay power spectra in the $k_{\perp} - k_{\parallel}$ plane related to the same LST bin. We found that the smooth diffuse foreground in the $k_{\perp} - k_{\parallel}$ plane dominates at low $k_{\parallel}$, with most power localized within $k_{\parallel} \le 0.2$~h~cMpc$^{-1}$. The foreground 
power drops by four to five orders of magnitude in the $k_{\parallel} \ge 0.2$~h~cMpc$^{-1}$ region, where the EoR signal is expected to dominate over the foreground emission. 
We notice some signature of a wedge-like structure in $k$ space \citep{Datta10,Morales12}, although in the current HERA antenna lay-out we are mostly limited to short baselines and hence the foreground wedge is not clearly visible. This wedge line is defined by \citep{Liu14, Dillon15}:
\begin{equation}
k_{\parallel} = \left[\mathrm{sin}(\theta_{\mathrm{field}}) \frac{H_{0}D_{M}(z)E
(z)}{c \, (1+z)}\right]k_{\perp},
\label{eq:wedge}
\end{equation}
where $\theta_{\mathrm{field}}$ is the angular radius of the field of view. We also show on the figure the wedge line corresponding to the horizon limit ($\theta_{\mathrm{max}} = 90^{\circ}$). 

\section{Gaussian Process Regression and foreground characterization} 
\label{sec:gpr}

The delay power spectrum results show that the data is mostly dominated by foregrounds. GPR offers a way to model these foregrounds in a maximum likelihood way. In this section, we summarize the GPR formalism \citep[for a detailed review of how the method works see][]{Mertens18} and apply it to model foreground components in HERA-47 observations.

In this framework, the different components of 21-cm observations, such as the astrophysical foregrounds, mode-mixing contaminants, and the 21-cm signal, are modelled with a Gaussian Process. A Gaussian Process is the joint distribution of a collection of normally distributed random variables \citep{Rasmussen05, Gelman14}. The covariance matrix of this distribution is specified by a covariance function, which defines the covariance between pairs of observed data points (i.e., at different frequencies). The co-variance function ultimately determines the structure that the GP will be able to model (for example, here the smoothness of the foregrounds). 

The GPR process requires the choice of the model for the covariance function and a selection of the best-fit parameters of such a model (what we call the hyper-parameters). Model selection is done in a Bayesian sense by maximizing the marginal-likelihood, also called the evidence, which is the integral of the likelihood over the prior range, given the data. For a fixed model, standard gradient-based optimization or Monte Carlo Markov Chain (MCMC) methods can be adapted to determine the best-fit parameters of the covariance functions. We note here that currently we model the data only in the frequency axis and no baseline dependence has been introduced in the hyper-parameter optimization with GPR (i.e., there is no dependence on baseline length). This assumption is supported by Figure~\ref{fig:1d_I_ps_bagpr}, where we can see that the power spectrum is similar for different baseline lengths.

In the following equations, $\mathbf{d}$ represents the time-averaged visibilities within a given LST bin and we have not explicitly shown the time dependence of the data. Considering an observed data $\mathbf{d}$ and a GP co-variance model which includes a foreground term $K_{\mathrm{f}}$ and a residual term (noise and 21-cm signal) $K_{\mathrm{r}}$, the data co-variance can be expressed as, $K = K_{\mathrm{f}} + K_{\mathrm{r}}$. 
After GPR, we can retrieve the foreground part of the signal  $E(\mathbf{f_{fg}})$ which always refers to the total signal except for noise or 21-cm signal through basically a Wiener filter \citep{Wiener}:
\begin{equation}
E(\mathbf{f_{fg}}) = K_{\mathrm{f}} \left[K_{\mathrm{f}} + K_{\mathrm{r}} \right]^{-1} \mathbf{d}.
\end{equation}
In the GPR context this is referred to as the posterior mean matrix while 
\begin{equation}
\mathrm{cov}(\mathbf{f_{fg}}) = K_{\mathrm{f}} - K_{\mathrm{f}} \left[K_{\mathrm{f}} + K_{\mathrm{r}} \right]^{-1} K_{\mathrm{f}}   \label{eq:covfg}  
\end{equation}
is the posterior co-variance matrix. 

Assuming that the GP co-variance model is optimal and taking $\langle \mathbf{d} \mathbf{d}^{H}\rangle = K_{\mathrm{f}} + K_{\mathrm{r}}$, then $\langle E(\mathbf{f_{fg}})E(\mathbf{f_{fg}})^H \rangle =
K_{\mathrm{f}} - \mathrm{cov}(\mathbf{f_{fg}})$. This highlights that to obtain the expected co-variance model of the foregrounds, $K_{\mathrm{f}}$, directly from $E(\mathbf{f_{fg}})$, we need to un-bias the estimator using $\mathrm{cov}(\mathbf{f_{fg}})$. 
We implement a similar unbiasing for the delay power-spectra of the different foreground components by first taking the delay transform of $E(\mathbf{f_{fg}})$ and $\mathrm{cov}(\mathbf{f_{fg}})$ and then adding them in the power spectrum domain. We finally normalize by the observed cosmological volume to construct the delay power spectrum $P(k_{\perp}, k_{\parallel})$ in units of ${\rm K^{2} \, [h^{-3}\, cMpc^{3}}]$. More specifically, we calculate the covariance matrices by fitting the hyper-parameters to all the data, while the posterior mean is obtained for each time-averaged visibility (so the covariance calculated from $E(\mathbf{f_{fg}})$ is not necessarily the same as the initial $K_{\mathrm{f}}$).
In this paper we consider the power spectrum of the different foreground components. This implies calculating $E(\mathbf{f_{fg}})$ for each of the foreground components, where we replace $K_{\mathrm{f}}$ by the optimized co-variance of the corresponding foreground component (while keeping the term in square brackets, $\left[K_{\mathrm{f}} + K_{\mathrm{r}} \right]$, the same, since it is the total co-variance).

\subsection{Covariance functions}
\label{sec:covfunc}
\begin{figure}
\includegraphics[scale=0.99]{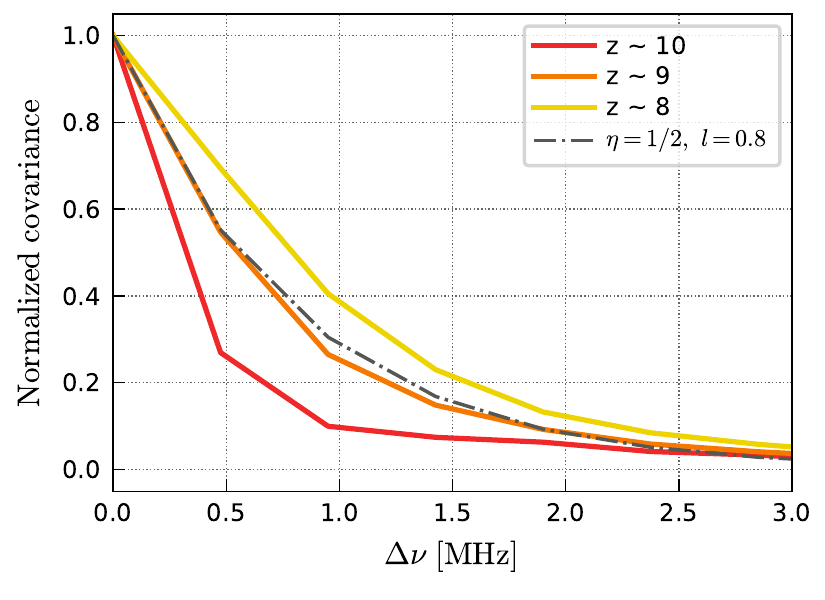} 
%\plotone{eor_21cmfast_co-variance_l08.pdf}
\caption{Example of the normalized GP exponential co-variance function with a frequency coherence-scale $l_{\mathrm{inj}} = 0.8$~MHz (dot-dashed line), compared to the co-variance of a simulated 21-cm EoR signal at different redshifts using 21cmFAST \citep{Mesinger07, Mesinger11}.}
\label{fig:GP_eor}
\end{figure}

In this section, we review the co-variance functions for the different components of the data. The selection of a co-variance function $\kappa$ for the 21-cm signal can be chosen by comparison to a range of 21-cm signal simulations. For this analysis, we choose a Matern $\eta = 1 / 2$ co-variance function with a frequency coherence-scale $l$ parameter: 
\begin{equation}
\label{eq:matern_cov}
\kappa_{\mathrm{Matern}}(\nu_p, \nu_q) = \sigma_{\rm f}^2 \frac{2^{1 - \eta}}{\Gamma(\eta)}\left(\frac{\sqrt{2\eta}r}{l}\right)^{\eta}
K_{\eta}\left(\frac{\sqrt{2\eta}r}{l}\right),
\end{equation}
where $\sigma_{\rm f}$ is the signal variance, $r = |\nu_q - \nu_p|$ and $K_{\eta}$ is the modified Bessel function of the second kind. The parameter $\eta$ controls the smoothness of the resulting function.
For $\eta = 1 /2$, the Matern kernel is equivalent to an exponential kernel. The choice of this co-variance kernel well matches the co-variance of the EoR signal with 21cmFAST \citep[][Figure~\ref{fig:GP_eor}]{Mesinger11}. Following \cite{Mertens18}, we used a uniform prior in the $0.01 - 1.25$~MHz range on the hyper-parameter $l$.

The intrinsic smooth foregrounds are modelled with a Radial Basis Function or RBF kernel (also known as the ``squared-exponential"  or a ``Gaussian''  kernel):
\begin{equation}
\label{eq:matern_rbf}
\kappa_{\mathrm{RBF}}(\nu_p, \nu_q) = \sigma_{\rm f}^2 \exp(- \frac{r^2}{2 l^2}),
\end{equation}
where the coherence scale $l$ controls the smoothness of the function, $\sigma_{\rm f}$ is the signal variance and the frequency coherence scale was bounded in the $10 - 200$~MHz range. We note that the Matern kernel (equation~\ref{eq:matern_cov}) is a generalization of the RBF kernel, parameterized by an additional parameter $\eta$. When $\eta$ tends to %$\infty$ 
infinity, the kernel becomes equivalent to RBF kernel.
Medium-scale fluctuations coming from a combination of the instrumental chromaticity and imperfect calibration (termed as `mode-mixing' components) are also modelled by a GP with an RBF covariance function where the characteristic coherence-scale $l$ is bounded in the $2 - 20$~MHz range.

\subsection{Foreground modelling}
\label{sec:modremFg}
Here, we discuss the GPR model foreground components, including the modeling and subsequent removal of the frequency and amplitude, modulated periodic signal with an additional GP co-variance kernel. Again, following \cite{Mertens18}, we modelled the GP
co-variance function by decomposing the foreground co-variance as:
\begin{equation}
\label{eq:fgcov}
K_{\mathrm{fg}} = K_{\mathrm{sky}} + K_{\mathrm{mix}}, 
\end{equation}
where the `sky' denotes the intrinsic smooth foreground sky and `mix' denotes the mode-mixing
contaminants which introduce oscillations in frequency mostly caused by the instrument. It is expected that $K_{\mathrm{sky}}$ will pick up the frequency dependence of the foreground signal at a given $uv$ point, whereas the mixing component $K_{\mathrm{mix}}$ can model relatively rapidly varying foreground components such as the fact that the $uv$ point itself also moves in the $uv$ plane with frequency \citep{Morales12}
and hence is sensitive to extra angular and frequency scale structures. 
We remind the reader that here we use the RBF kernel to model the foregrounds and an exponential kernel is used to represent the 21-cm signal co-variance function. To select the optimal mode-mixing co-variance function, we considered the Matern kernel with $\eta = 3/2$ and 5/2 along with the RBF kernel with a uniform prior in the $2-20$~MHz range. We found that the difference in the log-likelihood for the RBF kernel from the Matern 3/2 and the Matern 5/2 kernels are 1717 and 854 respectively (keeping the other covariances fixed). Based on this evidence, we choose to use the RBF kernel to model the mode-mixing component. 
%Similar tests can also be tried to choose the most appropriate kernel for other foreground model components. 
We optimize the
log-marginal-likelihood for the full set of visibilities (real and imaginary part separately) for the six variances and the coherence length scales hyper-parameters (namely, $\sigma_{21}^2$, $l_{21}$, $\sigma_{{\rm sky}}^2$, $l_{{\rm sky}}$, $\sigma_{{\rm mix}}^2$ and $l_{{\rm mix}}$), assuming the coherence scale is spatially invariant i.e. the same for each baseline type. The python package \textsc{GPy}\footnote{\url{https://sheffieldml.github.io/GPy/}} is used to do the optimization
using the full set of visibilities. The noise term is modelled with a fixed variance where the covariance matrix describes the variance along the frequency direction. The noise in the real data has both a frequency and a time dependence but here we choose only the frequency axis to approximate the noise variance. 
We found that the frequency coherence-scale of the `sky' and `mix' co-variance kernel are about 20~MHz and 2.4~MHz respectively. 

\begin{figure}
\includegraphics[scale=0.47]{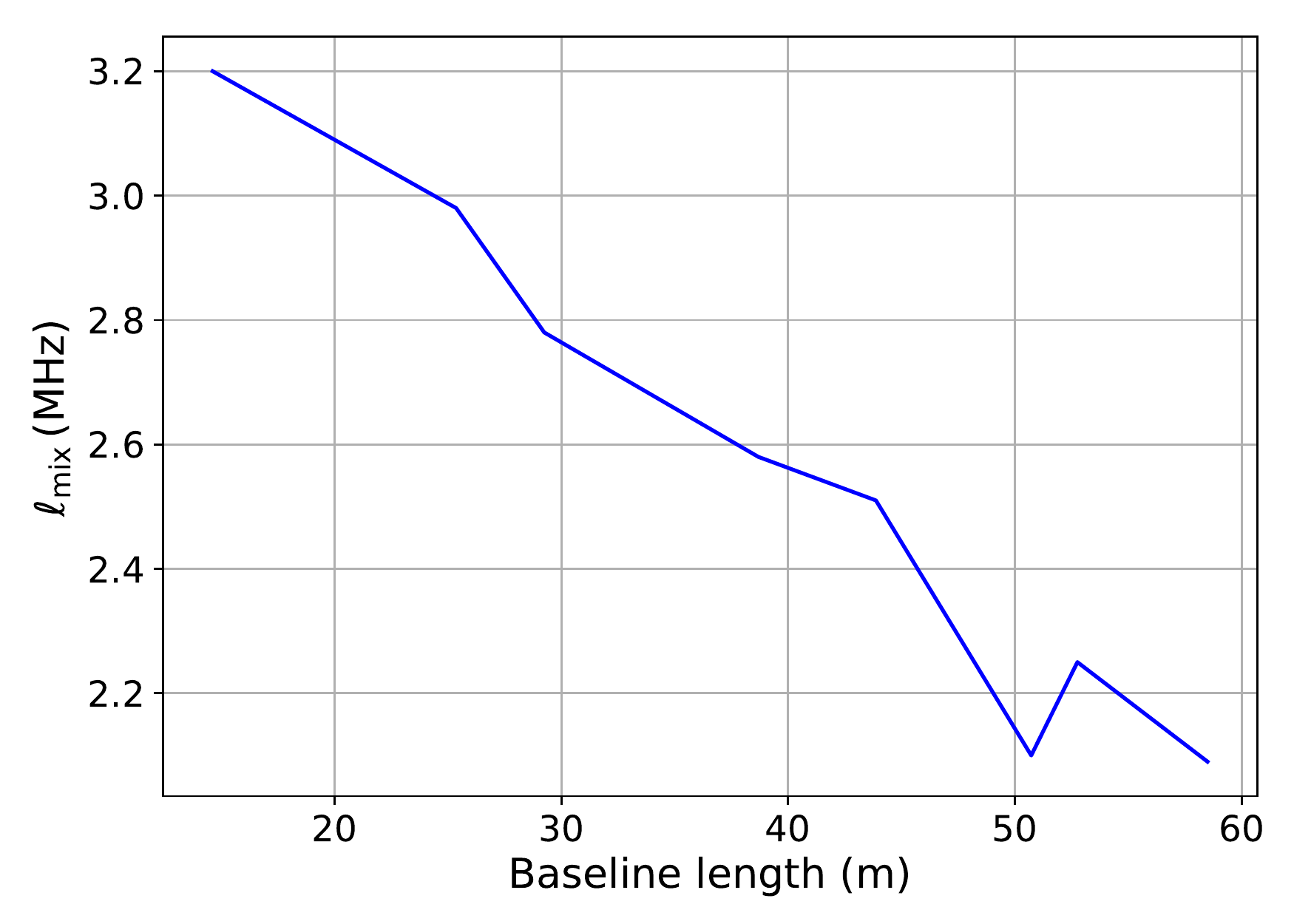}
%\plotone{Optimal_coh_base.png}
\caption{The optimized co-variance scale $l_{{\rm mix}}$ for the mode-mixing kernel as a function of baseline lengths using a `per-baseline' GPR technique.}
\label{fig:perbase-gpr-coh}
\end{figure}

In general the coherence scale is expected to be dependent on the baseline length, with longer baselines de-correlating faster than shorter baselines. 
We investigate this effect by implementing a `per-baseline' GPR approach which allows us to model a coherence scale for different baseline lengths using a smaller data-set with an increased number of degrees of freedom. 
We found that the coherence-scale decreases at longer baselines (Figure~\ref{fig:perbase-gpr-coh}), from $l \sim 3.2$~MHz for the 14.6~m baseline to $l \sim 2.2$~MHz for $\sim 60$~m baseline. In the `all-baselines' GPR implementation, the optimal coherence-scale was about 2.4~MHz, which falls inside the maximum-minimum range of `per-baseline' GPR approach. These results further agree well with our initial choice of the $2-20$~MHz prior range for the medium-scale frequency fluctuations introduced in Section~\ref{sec:covfunc}. The main reason for this behavior is the limited baseline range used in this analysis over which the foreground co-variance remains similar.

The inclusion of significantly longer baselines would likely require a `per-baseline' GPR fit as the foreground coherence will change more significantly across the range of baseline lengths. This could be implemented without significantly increasing the number of degrees of freedom by allowing the coherence-scale parameters to be a function of the baseline length. We leave this investigation for future work. 

We then considered all nights, coherently LST combined data sets for which we estimate different foreground components using GPR. For the GPR foreground modeling and the power spectrum estimation we used the python package \textsc{ ps\_eor}\footnote{\url{https://gitlab.com/flomertens/ps\_eor}}.

\begin{figure*}
\includegraphics[scale=0.45]{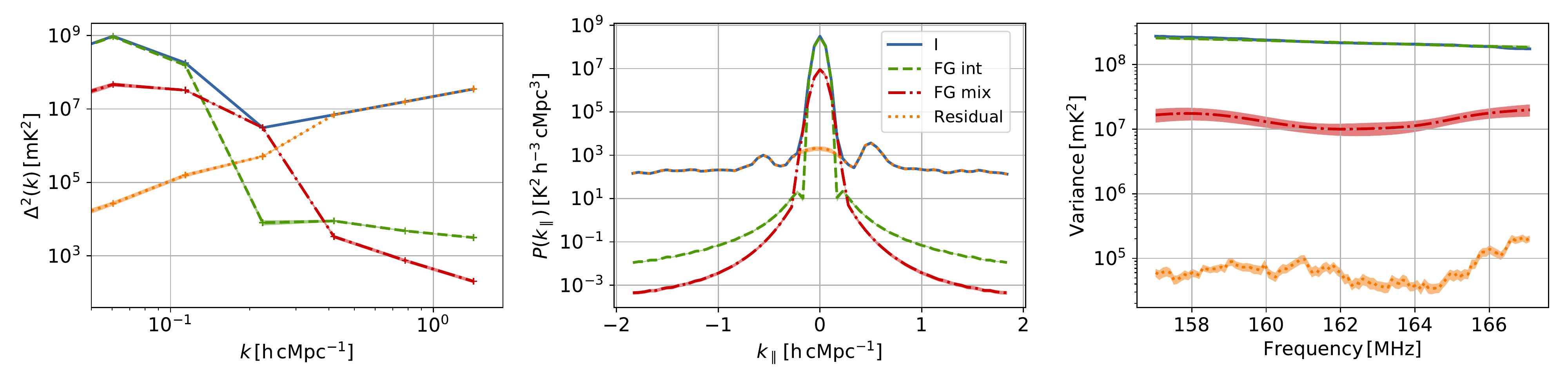}
\caption{Power spectra and derived foreground components after GPR analysis of the coherently averaged combined data: spherically averaged power spectra (left panel), delay spectra averaged over all baselines (middle panel). The right panel shows the variance of the different foreground components across frequency. The shaded area highlights the uncertainty ($2\sigma$) 
from the GP process (equation~\ref{eq:covfg}) and the uncertainty on the model fit from the MCMC run onto the foreground power spectrum components.}
\label{fig:gpr_fg_comp}
\end{figure*}

Figure~\ref{fig:gpr_fg_comp} shows the power spectrum and the variance across frequency for the different foreground components that we recover using the GPR technique. Note that in our GPR power spectrum estimation method, the hyper-parameters of the covariance model are optimized using the Bayesian evidence. Doing an MCMC we then get the posterior distribution of these hyper-parameters, and this is the first source of uncertainty that we use and propagate to the power spectra. The shaded area in Figure \ref{fig:gpr_fg_comp} shows the propagated $2\sigma$ uncertainty on the hyper-parameters and the uncertainty on the model fit (equation~\ref{eq:covfg}) and from the MCMC run (for details see Section~\ref{sec:FgparaMCMC}) onto the different foreground power spectrum components. An important point to note here is that the uncertainty on the power spectra that we estimate are correct assuming that our assumed co-variance functions are appropriate.

We notice that the `FG mix' model has a small coherence scale ($2 - 3$~MHz) and therefore the variance has a wave-like pattern, but for the intrinsic foregrounds it is mostly smooth across frequency. We detect a `bump' in the power spectrum around $k_{\parallel} \sim 0.5$~h~cMpc$^{-1}$, corresponding to a $\sim 900$~ns delay, indicating the presence of a non-negligible contamination in the data (possibly due to internal signal chain reflections, or a more dominant instrumental cross-talk feature spanning delays of 800 - 1200~ns which does not look like an EoR signal and can therefore be filtered out \citep{Kern19b}) which we investigate in more detail in Section~\ref{sec:per_sig}.

\begin{figure}
\includegraphics[scale=0.55]{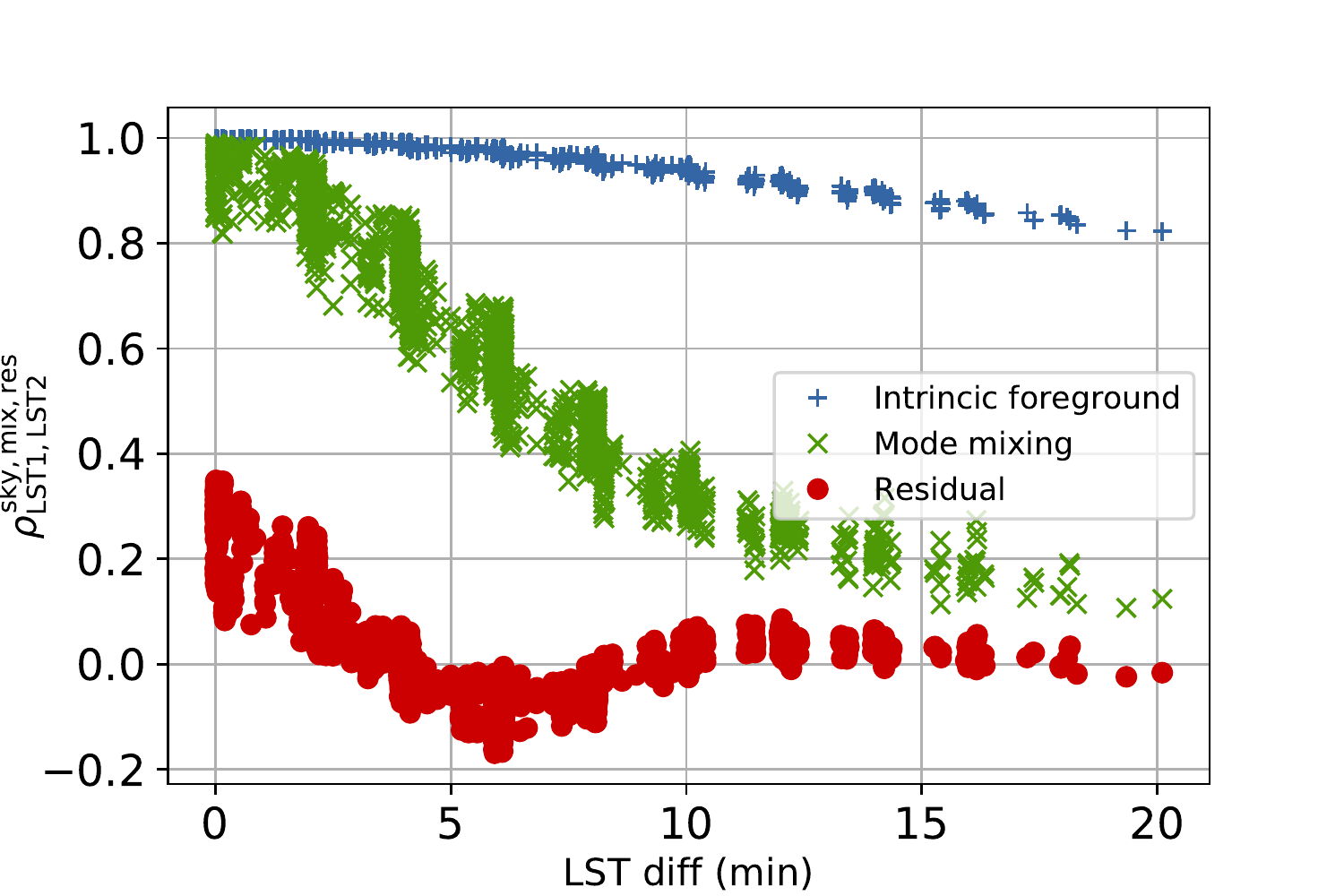}
\caption{Correlation coefficient $\rho_{{\rm LST1, LST2}}^{{\rm sky, mix, res}}$ as a function of LST difference for all the night, LST-binned data. Each marker symbol (`+', `x' or `o') represents an LST difference.} 
\label{fig:gpr_fg_corrl_lst}
\end{figure}

Figure~\ref{fig:gpr_fg_corrl_lst} shows the correlation along the frequency direction of the different GPR components (intrinsic foreground, mode-mixing foreground and residuals) as a function of LST difference for all the combinations of LST binned data sets, covering a LST range of $\sim 20.92^{\rm h}-21.26^{\rm h}$ (each cross represents an LST bin difference). Here, we compute the correlation for every combination of the LST binned visibility data sets for each baseline and finally we average over the baselines to determine the final value. The correlation coefficient is given by:
\begin{equation}
\label{eq:corrl_coeff}
\rho_{{\rm LST1, LST2}}^{{\rm cmpt}} = \frac{{\rm cov\left({\rm fg_{LST1}^{{\rm cmpt}}, \rm fg_{LST2}^{{\rm cmpt}}}\right)}}{\sigma\left(\rm fg_{LST1}^{{\rm cmpt}}\right) \sigma\left(\rm fg_{LST2}^{{\rm cmpt}}\right)}
\end{equation}
where, $\rm fg_{LST1}^{{\rm cmpt}}$ and $\rm fg_{LST2}^{{\rm cmpt}}$ are the foreground model components corresponding to the `sky', `mix' or the residual at two different LSTs.
For large LST differences, the correlation should go down since we are looking at different parts of the sky. From Figure~\ref{fig:gpr_fg_corrl_lst} we see that the intrinsic foreground correlation remains above $80 \%$ regardless of the time difference. The correlation coefficient starts to decay only for LST differences $> 2-4$ min (as the sky starts to shift). The mode-mixing de-correlates significantly as a function of LST difference. This typically depends on the coherence scale in the $uv$-plane as a baseline moves through it and is faster for longer baselines. The mode-mixing is also more affected by LST difference de-correlation because it contains fluctuations due to small beam differences mainly further away from the phase center.
%This de-correlation is generally faster at longer baselines as, at different LSTs, the combination of a different sky with the beam produces different mode-mixing signatures. 

\subsubsection{Characteristics of the periodic signal}
\label{sec:per_sig}
\begin{figure}
\includegraphics[scale=0.55]{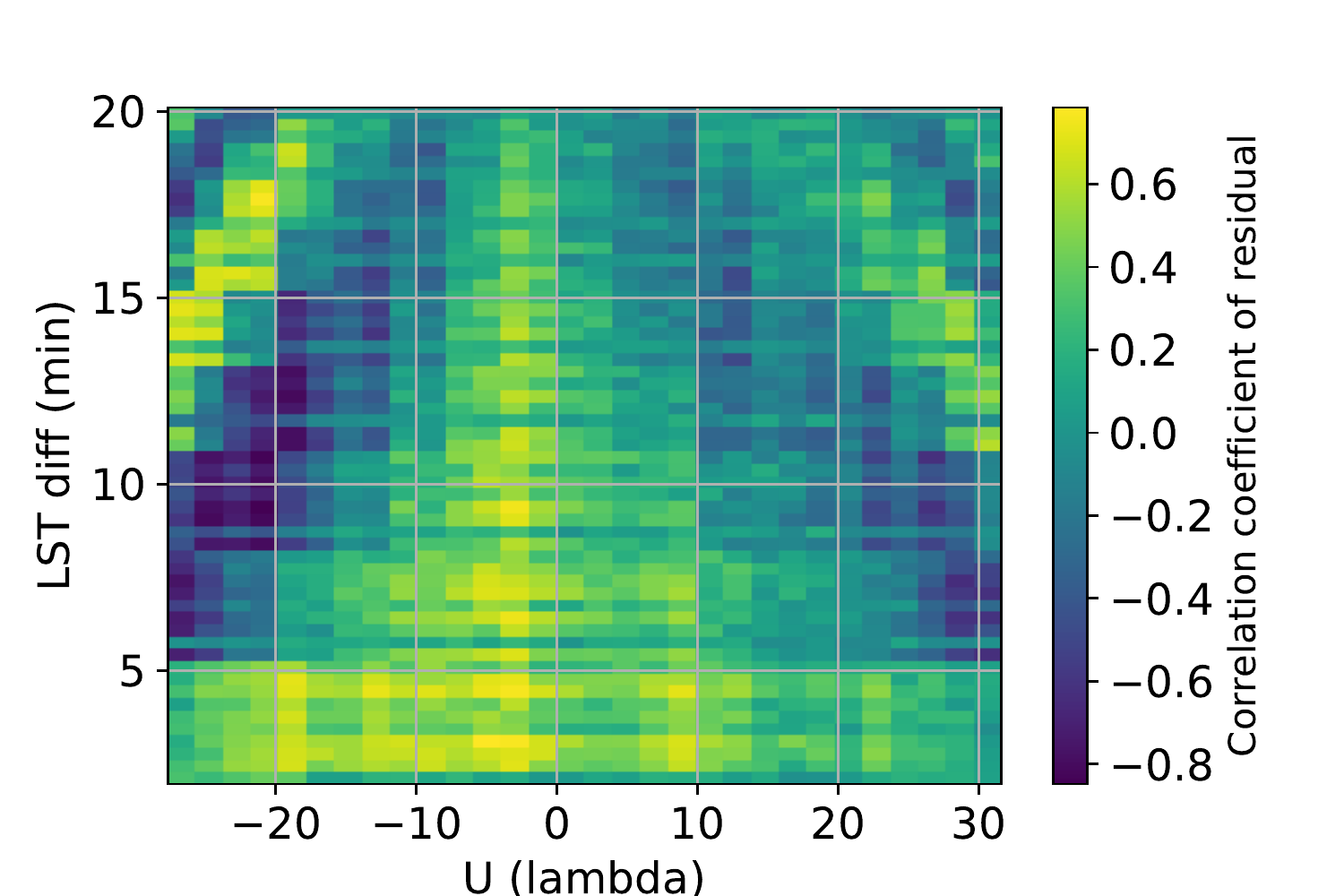}
%\plotone{Corrl_U_LST_diff.pdf}
\caption{Correlation coefficients of the residual visibility data after GPR as a function of baseline length and LST difference. The plot shows very strong dependence of the periodic signal on baseline length $|{\bf u}|$ and it also varies with LST difference.}
\label{fig:corrl_basl_lstdiff}
\end{figure}

\begin{figure}
\includegraphics[scale=0.55]{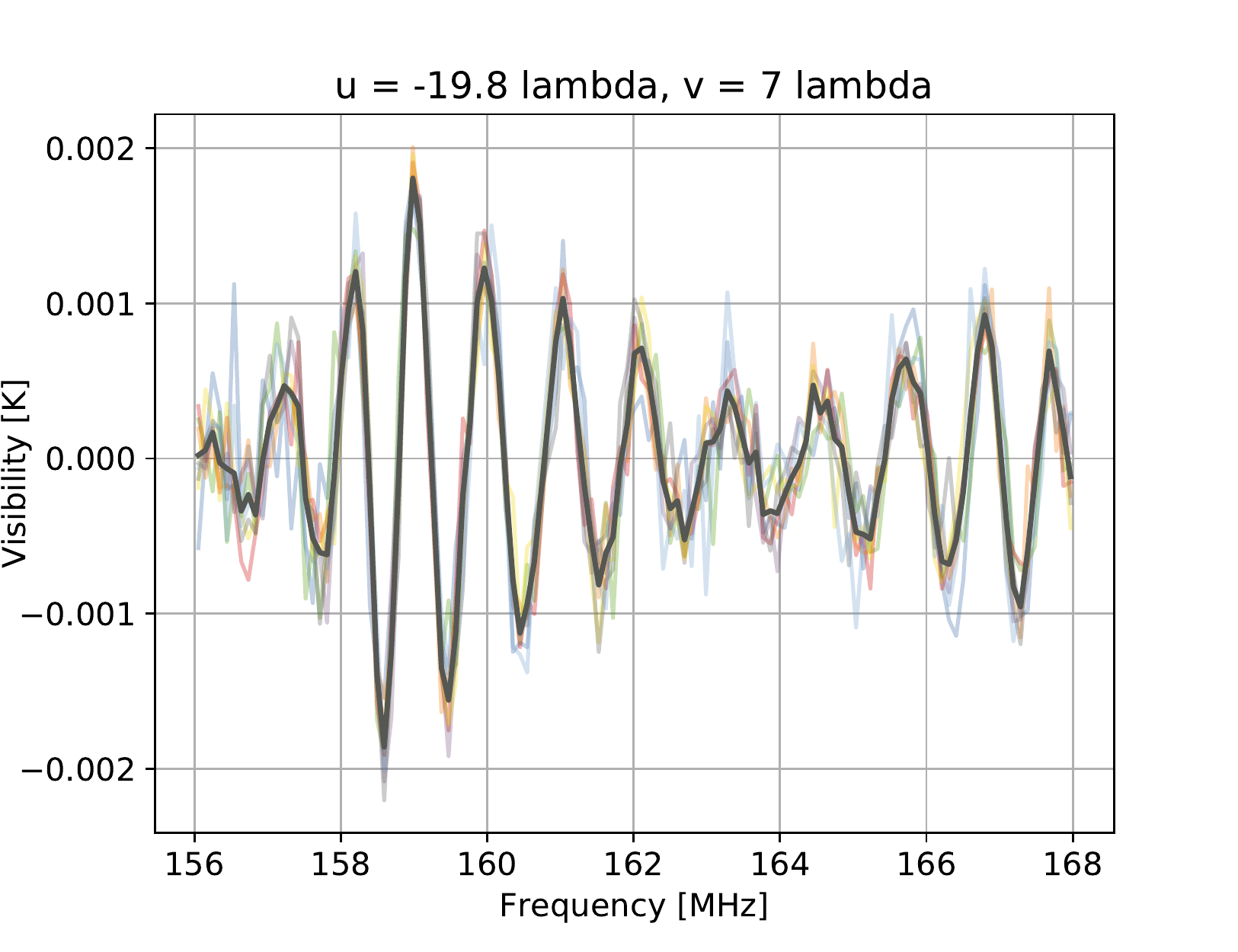}
%\plotone{per_signal_visibility_fit.pdf}
\caption{Example of the periodic signal for 38.6~m baseline with coordinates $(u, v) =(-19.8, 7)\,\lambda$. The real part of the visibility is shown in `K' units. The different transparent lines correspond to different LST data sets on which we apply a frequency phase offset to align the periodic signal. This signal can be fitted using GPR with a combination of an RBF and Cosine co-variance kernel (solid line).}
\label{fig:per_signal_visibility_fit}
\end{figure}

After foreground removal, the residual power spectrum is dominated by noise and an almost periodic signal which reveals itself by an excess power at $k_{\parallel} \sim 0.5$~h~cMpc$^{-1}$. We find the periodic signal is baseline dependent and it also varies with LST difference. Figure~\ref{fig:corrl_basl_lstdiff} shows the correlation of the residual visibilities (equation~\ref{eq:corrl_coeff}) for different baselines as a function of LST difference. Two periodic signals from two different LST times appear to be phase shifted. A closer inspection reveals that the amplitude and periodicity of this signal does not remain stationary but varies with frequency. For example, the residual visibilities for a specific baseline and the fit to the periodic signal is shown in Figure~\ref{fig:per_signal_visibility_fit}. Similar frequency-dependent complex patterns are also seen for other baselines. 
This profile can be fitted using the GPR method and a combination of a RBF and Cosine co-variance function, $K_{\mathrm{per}}$, on each baseline individually. The co-variance function $\kappa_{\mathrm{per}}$ for the periodic kernel depends on the characteristic coherence-scale $l_{\mathrm{per}}$ over which the periodic signal vary, the signal variance $\sigma_{\mathrm{per}}^2$, and the period $p_{\mathrm{per}}$:
\begin{equation}
\label{eq:per}
\kappa_{\mathrm{per}}(\nu_p, \nu_q) = \sigma_{\rm per}^2 \, \exp{\left(- \frac{r^2}{2 l_{\rm per}^2}\right)} \cos\left(\frac{2 \pi r}{p_{per}}\right).
\end{equation}
We found the main periodicity is $\sim 1$~MHz.

\citet{Kern19b} provides a thorough investigation of such systematic effect, attributing it to a combination of instrumental cross-coupling (e.g. mutual coupling and crosstalk) and cable reflections within the analog signal chain.
\citet{Kern19a} present methods for modeling and removing these systematic terms in the data.
In the following section, we show how it can be modeled and subtracted in the GPR formalism.

\subsubsection{Filtering the periodic signal with GPR}
\label{sec:FgperGPR}
 To model this locally periodic signal with amplitude varying over a certain coherence-scale, we introduce an additional kernel in the foreground co-variance model. We use a combination of a RBF and a cosine kernel to model the period in frequency. The updated foreground co-variance function is modelled as:
\begin{equation}
\label{eq:upfgcov}
K_{\mathrm{fg}} = K_{\mathrm{sky}} + K_{\mathrm{mix}} + K_{\mathrm{per}} 
\end{equation}
where, the $K_{\mathrm{per}}$ represents the periodic signal contaminant (see equation~\ref{eq:per}). We used this updated foreground co-variance model in our GP optimization. The  GPR estimates of the parameters for the periodic co-variance function are found to be $p_{\mathrm{per}} \sim 1$~MHz and $l_{\mathrm{per}} \sim 1.2$~MHz respectively.

\begin{figure*}
\includegraphics[scale=0.45]{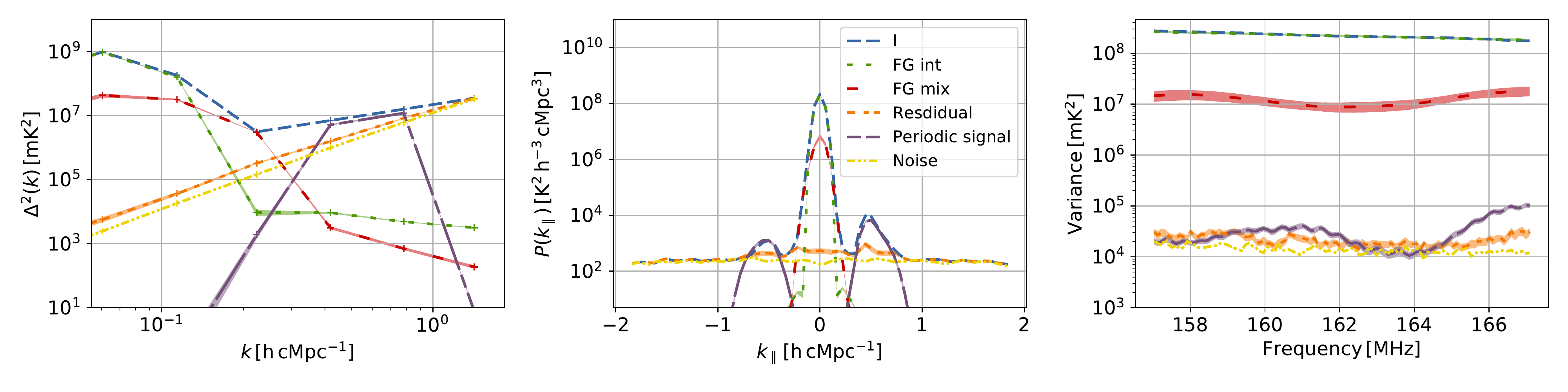}
%\plotone{GPR_fg_gprfilper_comp.pdf}
\caption{Same as Figure~\ref{fig:gpr_fg_comp} but now including the periodic signal and the estimated noise power spectrum and variance.} 
\label{fig:gpr_fg_periodic_comp}
\end{figure*}

Figure~\ref{fig:gpr_fg_periodic_comp} displays the power spectrum of different GPR modelled components including the periodic signal. 
We notice that the periodic signal peaks around $\sim 0.4 - 0.8$~h~cMpc$^{-1}$. 
%where it becomes comparable to the Stokes I signal. 
In the middle panel, we display the GPR model that nicely isolates the periodic signal component around $k_{\parallel} \sim 0.5$~h~cMpc$^{-1}$. In general, we find that the periodic signal is $k$-dependent. It appears at $k \sim 0.17$~h~cMpc$^{-1}$, reaching a $\sim 10^7$~mK$^2$ peak at $k \sim 0.4$~h~cMpc$^{-1}$ - approximately six orders of magnitude brighter than the expected 21-cm power spectrum \citep{Mesinger11}. 

The average variance across frequency for the periodic signal component is $\sim 3.8 \times 10^4$~mK$^2$, while the mean-variance of the mode-mixing signal is around $\sim 1.2 \times 10^7$~mK$^2$, approximately three orders of magnitude higher. 
The noise power spectrum shown in Figure~\ref{fig:gpr_fg_periodic_comp} is estimated by splitting the data set in even and odd times with a 10.7~s time separation and taking the difference between the two. At this time resolution, the foregrounds cancel out almost perfectly. We find the residual power spectrum level is close to the estimated noise power spectrum, especially at $|k_{\parallel}| \ge 0.85$~h~cMpc$^{-1}$.

\begin{figure}
\includegraphics[scale=0.55]{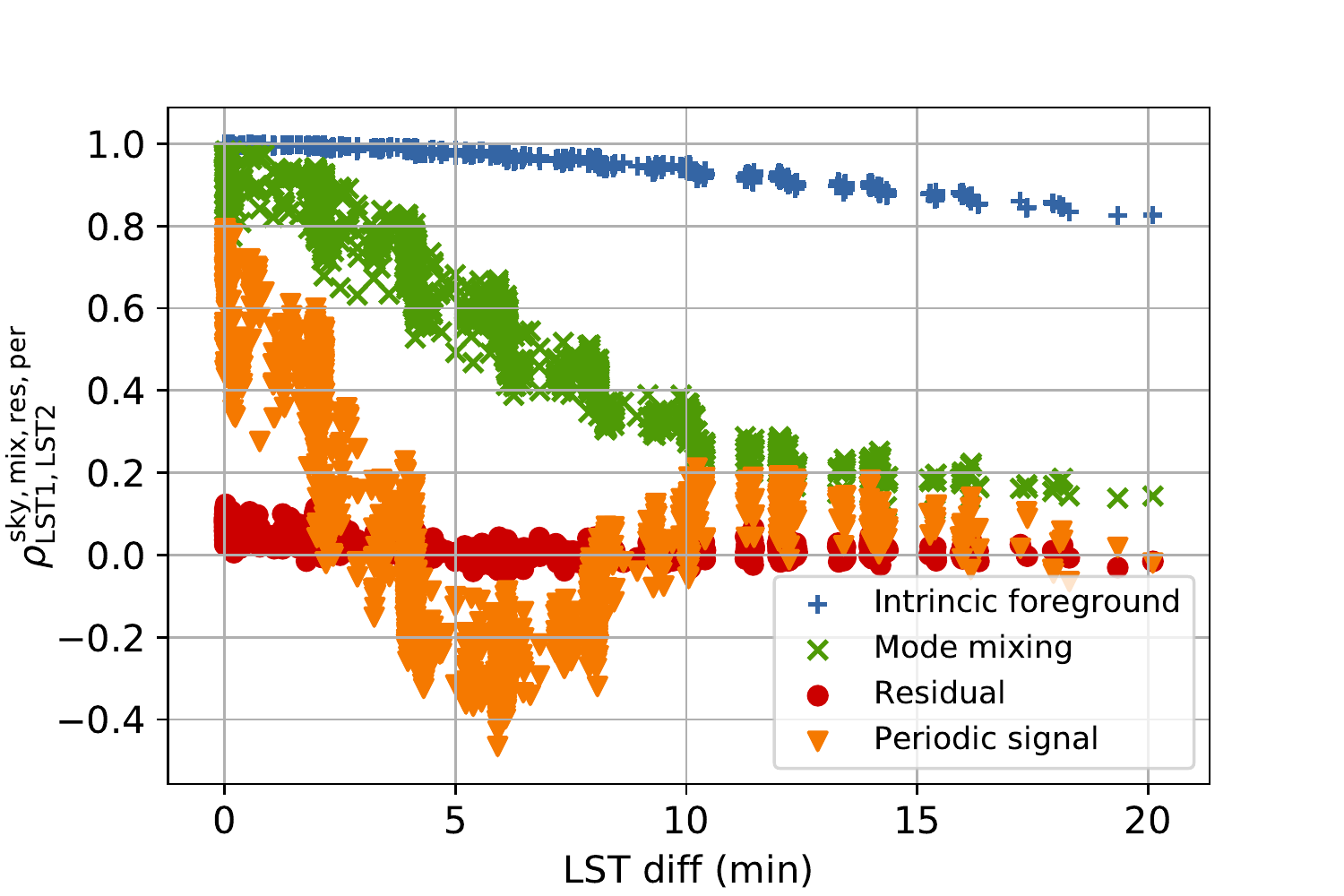}
%\plotone{GPR_fg_gprfilper_corrl_LST.pdf}
\caption{Same as Figure~\ref{fig:gpr_fg_corrl_lst}, but here the periodic signal has been modelled with a RBF and cosine kernel and then removed.}
\label{fig:gpr_fg_corrl_lst_gprfilper}
\end{figure}

The residual in Figure~\ref{fig:gpr_fg_corrl_lst_gprfilper} reveals that there is still some time correlation left, but overall, we find the residuals have become more uncorrelated and noise-like compared to Figure~\ref{fig:gpr_fg_corrl_lst} where the GP foreground co-variance was modelled only with a combination of `sky' and `mode-mixing' kernels.

\subsubsection{Foreground model hyper-parameter uncertainties}
\label{sec:FgparaMCMC}

We sampled the posterior distribution of the foreground model hyper-parameters and characterize their correlation with a MCMC. We used the \textsc{emcee} python package\footnote{\url{http://dfm.io/emcee/current/}}~\citep{Foreman-Mackey13} which uses an ensemble sampler algorithm based
on the affine-invariant sampling algorithm~\citep{Goodman10}. 

\begin{figure*}
\includegraphics[scale=0.30]{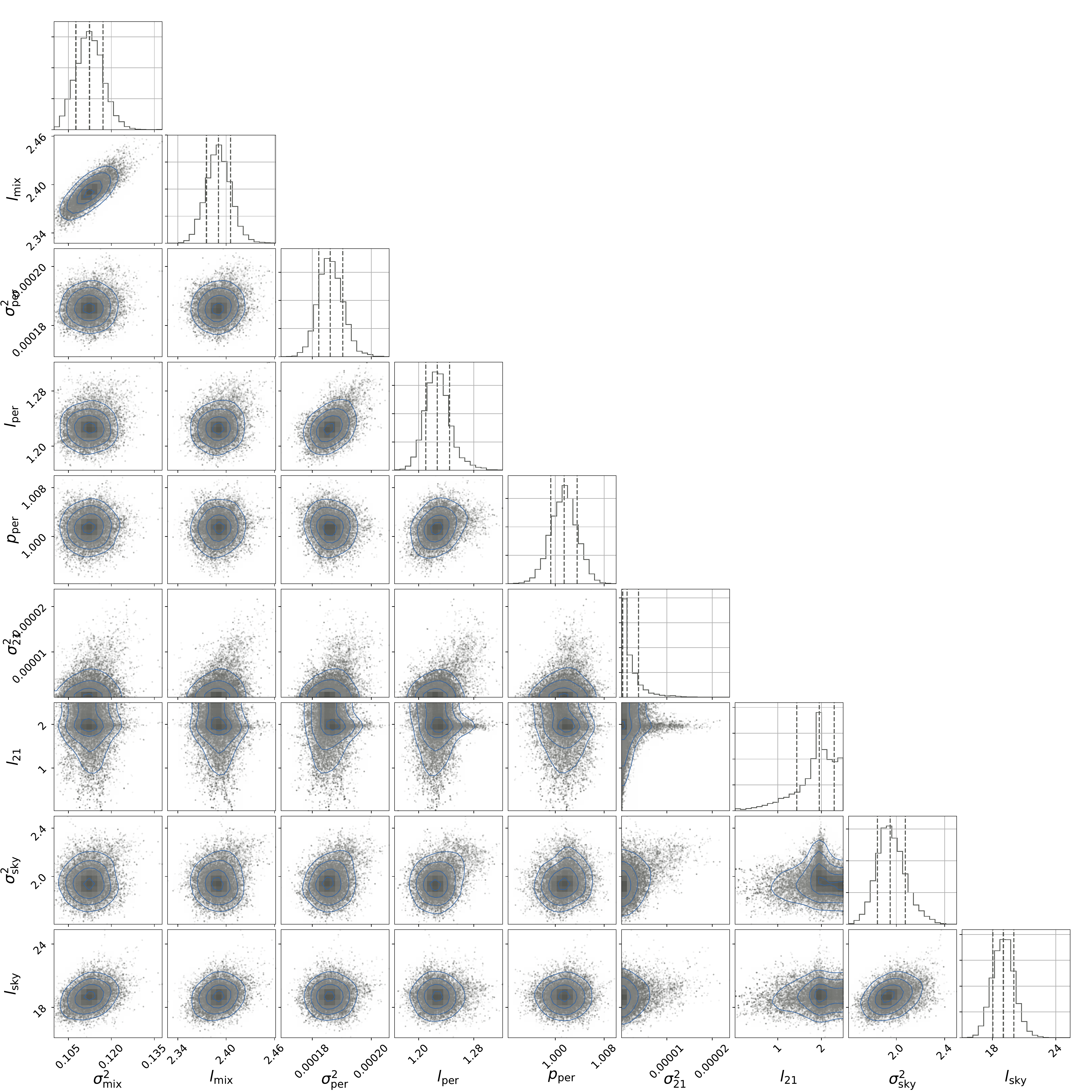}
\caption{Posterior probability distributions of the GP model hyper-parameters. We show here the coherence-scale and strength of the EoR co-variance kernel ($l_{21}$ in MHz and $\sigma_{21}$ in ${\rm K}^2$), the coherence-scale and strength of the mode-mixing foreground kernel ($l_{{\rm mix}}$ in MHz, $\sigma_{{\rm mix}}$ in ${\rm K}^2$), the intrinsic foreground kernel ($l_{{\rm sky}}$ in MHz, $\sigma_{{\rm sky}}$ in ${\rm K}^2$) and the periodic co-variance kernel hyper-parameters ($l_{{\rm per}}, \, p_{{\rm per}}$ in MHz and $\sigma_{{\rm per}}^2$ in ${\rm K}^2$). The vertical dashed lines show the first, second and third (Q1, Q2 and Q3) quantile levels. The diagonal panels represent the marginalized probability distribution of each parameter.}
\label{fig:post_prob_mcmc_per_9para}
\end{figure*}

Figure~\ref{fig:post_prob_mcmc_per_9para} shows the resulting posterior probability distribution of the GP model hyper-parameters. The variance of the EoR kernel, which was modelled with a GP exponential kernel, is found to be un-constrained and low. The data can be well modelled by the `sky', `mix', `per' (periodic) foreground kernels and the noise covariance matrix (modelled with a fixed variance) which contributes a large part of the variance at large $k_{\parallel}$. We compared the evidence values with and without the EoR co-variance kernel in the GP optimization. We find the evidence remains mostly unchanged and the Bayes factor \citep{Jeffreys61} is around $\sim 0.93$ for GP models with and without the EoR co-variance kernel. This essentially confirms that the signal is dominated by a noise-like component once the foregrounds are removed and adding an EoR kernel has an insignificant effect. Overall the confidence intervals of other kernel hyper-parameters are reasonably well constrained, except the variance of
the `21-cm signal' component which is consistent with zero. The significance of the coherence-scale of the `21-cm signal' is also reduced given the non-significant variance of this component. Table~\ref{tab:post_prob_mcmc_9para} highlights the parameter estimates and confidence intervals for the posterior probability distribution of the foreground model hyper-parameters. The estimated median values of the 
frequency coherence-scale of the `sky' and `mix' covariance kernel is about 19.4~MHz and 2.4~MHz respectively which is close to the GPR optimized values as presented in Section~\ref{sec:modremFg}.

\begin{table}
    \centering
    \caption{Summary of the estimated median and confidence intervals (first and third quantile levels (Q1 and Q3)) of
    the respective GP model hyper-parameters including the periodic co-variance kernel. \label{tab:post_prob_mcmc_9para}}
    \begin{tabular}{ccc}
        \hline
        Hyper-parameter & Prior & Estimate\\
        \hline
        %$l_{21}$ & $\mathcal{U}(0.001, 2.5)$ & $1.71 \substack{+0.46\\ -0.55}$  \\
        %$\sigma_{21}^2$ & $\mathcal{U}(1e-20, 0.1)$ & $0.000002 \substack{+0.000003\\ -0.000001}$  \\
        %$\sigma_{21}^2$ & $\mathcal{U}(1e-20, 0.1)$ & $-$  \\
        %\hline
        $l_{{\rm mix}} \ ({\rm MHz})$ & $\mathcal{U}(2, 20)$ & $2.40 \substack{+0.02\\ -0.01}$  \\
        $\sigma_{{\rm mix}}^2 \ ({\rm K}^2)$ & $\mathcal{U}(0.1, 0.9)$ & $0.115 \substack{+0.007\\ -0.005}$  \\
        \hline
        $l_{{\rm sky}} \ ({\rm MHz})$ & $\mathcal{U}(10, 200)$ & $19.42 \substack{+1.25\\ -1.18}$  \\
        $\sigma_{{\rm sky}}^2 \ ({\rm K}^2)$ & $\mathcal{U}(0.02, 2.5)$ & $1.89 \substack{+0.10\\ -0.09}$  \\
        \hline
        $l_{{\rm per}} \ ({\rm MHz})$ & $\mathcal{U}(1, 5)$ & $1.23 \substack{+0.01\\ -0.01}$  \\
        $p_{{\rm per}} \ ({\rm MHz})$ & $\mathcal{U}(0.628, 1.256)$ & $0.999 \substack{+0.002\\ -0.002}$  \\
        $\sigma_{{\rm per}}^2 \ ({\rm K}^2)$ & $\mathcal{U}(0.00001, 0.01)$ & $0.000183\substack{+0.000004\\ -0.000003}$  \\
    \end{tabular}
\end{table}

\section{Discussion and Conclusions}
\label{sec:discussion}

In this paper, we have used a novel foreground separation method, first introduced in \citet{Mertens18}, in order to model foregrounds with the HERA-47 array. The mainstream HERA data analysis takes advantage of the concept of avoiding foregrounds and provides a single-baseline power spectrum estimate: recent data analysis showed evidence of systematic effects that contaminate the EoR window, and motivated the development of strategies to mitigate their impact to the avoidance paradigm \citep{Kern19c}. An alternative effort to detect the 21~cm signal using closure quantities is actively being pursued \citep{Thyagarajan18,Carilli18,Carilli20}.

The method presented here uses Gaussian Process Regression to model various stochastic foreground components, such as the spectrally smooth intrinsic sky, mode-mixing components generating from the chromatic instrument and imperfect calibration, as well as a 21-cm signal. It therefore bears analogies with the avoidance approach as they both attempt to model and subtract systematic effects in the EoR window, but also more broadly models the foreground emission - which is not within the purpose of the avoidance approach. Foreground modeling may be a necessary step in order to reduce the leakage in the EoR window and access the high signal-to-noise ratio small $k$ modes \citep[e.g.,][]{Kerrigan18,Ewall-Wice20,Lanman20}.
%Moreover, our method combines together baselines with different lengths, leading to a sensitivity increase with respect to the single-baseline power spectrum often used in the avoidance approach \citep[e.g.,][]{Kolopanis19}.

Our analysis included a different co-variance function for each of intrinsic sky, mode-mixing and 21-cm signal components in the GP modeling. 
We found that the frequency coherence-scale of the `sky' and `mix' co-variance kernel are about 20~MHz and 2.4~MHz respectively. As a comparison, the typical (theoretical) frequency coherence scale for the 21-cm EoR signal is found to be around $\sim 0.8$~MHz, when fitted to the co-variance of a simulated 21-cm EoR template. The foreground power spectrum is shown to be contaminated by a $\sim 1$~MHz periodic signal whose amplitude changes from baseline to baseline. The periodic signal dominates the $0.25 < k < 0.9$~h~cMpc$^{-1}$ range. We included a combination of RBF and a cosine kernel to model this signal within our GPR method and found a fairly cleaner and flatter residual power spectrum across the $0.05 < k < 1.83$~h~cMpc$^{-1}$ range. The residual power spectrum is also mostly consistent with the estimated noise power spectrum, especially at high $k_{\parallel}$ values, whereas residuals are still present in the foreground and periodic signal-dominated region of the pwoer spectrum.

%With foreground removal we can possibly access lower $k$ modes where we expect the 21-cm signal to be brighter. Moreover, foreground removal can suppress contaminants at low $k$ modes which can reduce foreground leakage in the EoR window and thus assist the foreground avoidance approach close to the foreground wedge boundary. 
As foreground subtraction is potentially at risk of altering the 21-cm signal, we plan to further explore this approach using more HERA data and test the cleaning with signal injection tests using full-scale HERA simulations. 
In this paper, we have restricted ourselves to foreground modeling only and left the characterization of residual power spectra to future work, which will include end-to-end signal injection tests. 

Finally, we note that the foreground model used in this paper might not be complete, although it seems to be enough at this noise level. In particular it does not include other foregrounds contaminants such as the instrumental polarization leakage, residual RFIs, and the phase errors caused by the ionosphere or imperfect calibration. We plan to include these additional subtle effects in our GP co-variance modeling. In addition to these, we intend to implement a per-baseline GPR approach where the coherence-scale parameters are a function of the baseline length without exploding the number of degrees of freedom of the GPR fit. This will be relevant for longer HERA baselines where the larger baselines will de-correlate faster compared to the shorter baselines. Also, the present mode-mixing model can be improved by integrating the $k_{\perp}$ dependency of the foreground wedge. We further plan to include the isotropic nature of the 21-cm signal and its evolution at different redshift bins which will also ensure a more sensitive and detailed modeling.

\section*{acknowledgments}
We thank the anonymous referee and the editors for their useful comments and suggestions. 
This material is based upon work supported by the National Science Foundation under Grant Nos. 1636646 and 1836019 and institutional support from the HERA collaboration partners. This work is funded in part by the Gordon and Betty Moore Foundation and the National Research Foundation of South Africa (grants No. 103424 and 113121). HERA is hosted by the South African Radio Astronomy Observatory (SARAO), which is a facility of the National Research Foundation, an agency of the Department of Science and Innovation. Parts of this research were supported by the Australian Research Council Centre of Excellence for All Sky Astrophysics in 3 Dimensions (ASTRO 3D), through project number CE170100013. AG would like to thank SARAO for support through SKA postdoctoral fellowship, 2016. AG would also like to thank Dr. Sushanta Kumar Mondal for editing some of the figures. FGM and LVEK would like to acknowledge support from a SKA-NL Roadmap grant from the Dutch ministry of OCW. LVEK, FGM, BG also acknowledge support by an NWO-NRF Exchange Programme in Astronomy and Enabling technologies in Astronomy (NWO grant number 629.003.021). GB acknowledges support from the Royal Society and the Newton Fund under grant NA150184. MGS acknowledges support from the South African Radio Astronomy Observatory and the National Research Foundation (Grant No.~84156). GB acknowledges funding from the INAF PRIN-SKA 2017 project 1.05.01.88.04 (FORECaST). We acknowledge the support from the Ministero degli Affari Esteri della Cooperazione Internazionale - Direzione Generale per la Promozione del Sistema Paese Progetto di Grande Rilevanza ZA18GR02. AL acknowledges support from the New Frontiers in Research Fund Exploration grant program, a Natural Sciences and Engineering Research Council of Canada (NSERC) Discovery Grant and a Discovery Launch Supplement, the Sloan Research Fellowship, as well as the Canadian Institute for Advanced Research (CIFAR) Azrieli Global Scholars program. We acknowledge the HERA staff who made these observations possible.

\bsp	% typesetting comment
\label{lastpage}
\end{document}